\documentclass[manuscript]{aastex}
\usepackage{graphicx}
\usepackage{amssymb}
\usepackage{lscape}
\usepackage{cleveref}
\begin{document}
\title{Fe I oscillator strengths for transitions from high-lying odd-parity levels}
\author{M. T. Belmonte, J. C. Pickering, M. P. Ruffoni}
\affil{Blackett Laboratory, Physics Department, Imperial College London, London SW7 2AZ, UK}
\email{m.belmonte-sainz-ezquerra@imperial.ac.uk}
\author{E .A. Den Hartog, J. E. Lawler and A. Guzman}
\affil{Department of Physics, University of Wisconsin-Madison, Madison, WI 53706.}
\author{U. Heiter}
\affil{Observational Astrophysics, Department of Physics and
Astronomy, Uppsala University, Box 516, 751 20 Uppsala, Sweden}

\begin{abstract}

We report new experimental Fe I oscillator strengths obtained by
combining measurements of branching fractions measured with a Fourier
Transform spectrometer and time-resolved laser-induced fluorescence
lifetimes. The study covers the spectral region ranging from 213 to 1033 nm.
A total of 120 experimental log$(gf)$-values coming from 15 odd-parity energy levels are provided, 22 of which have not been reported previously and 63 values with lower uncertainty than the existing data. Radiative lifetimes for 60 upper energy levels are presented, 39 of which have no previous measurements.
\end{abstract}


\section{Introduction}

Iron is one of the most studied elements within the field of
astronomy due to its important presence in stellar spectra. With a very complex spectrum, neutral iron presents thousands of transitions across a very wide spectral range, from the ultraviolet to the infrared.  A very comprehensive study of its spectrum was undertaken by \cite{ref:nave1994}, which provides an extremely useful guide for the identification of Fe I spectral lines in astronomical spectra. However, of the 6758 lines included in the National Institute of Standards and Technology (NIST) atomic database in the spectral interval ranging from 200 to 1050 nm, only a small percentage possess accurate values of transition probabilities.

Atomic oscillator strengths (transition probabilities, log$(gf)$)
are essential to model stellar line intensities and calculate not
only chemical abundances, but also other stellar parameters. In
particular, the iron spectrum is of the utmost importance to obtain
stellar metallicities, as this property is directly linked to the
iron abundance. However, the quantity and quality of the existing
data lies far from the current needs of the astronomical community,
remaining the Achilles' heel of the field of Galactic archaeology
\citep{ref:bigot2006}. Several attempts have been made to assemble
comprehensive line lists with reliable atomic data
\citep{ref:heiter2015} which can be used as a standard input for the
different LTE and non-LTE models used to determine chemical
abundances.

Several studies regarding the measurement of oscillator strengths of neutral iron have been conducted over the last fifty years. A very detailed review of the situation  was carried out ten years ago by \cite{ref:fuhr2006}, where they present the most comprehensive compilation of Fe I transition probabilities to date which states clearly the need for new studies that complete and improve the quality of thousands of spectral lines in the database of Fe I log$(gf)$ values. Amongst all the works whose values are included in \cite{ref:fuhr2006}, two deserve special attention due to the quality of their results and their coverage. These are the experiments conducted by Blackwell et al. (1979a, 1979b, 1980, 1982a, 1982b and 1986) and \cite{ref:obrian1991}.

Very accurate absorption oscillator strengths were obtained by Blackwell et al. from a very stable light source in an absorption experiment, with estimated uncertainties lower than $\pm$4\% on an absolute scale. Their values are generally considered as the most reliable ones by \cite{ref:fuhr2006}, who rated them as `A' in their compilation. The comprehensive work from \cite{ref:obrian1991} provides accurate transition probabilities for 1814 spectral lines of neutral iron obtained in an emission experiment by using two different methods.
One method combines radiative lifetimes of 186 energy levels with measurements of branching fractions yielding 1174 absolute transition probabilities. The other method used by O'Brian et al interpolated the populations of energy levels using those with known lifetimes in an inductively coupled plasma source, producing 640 extra transition probabilities with uncertainties that they estimated to be lower than $\pm$10\%. Within the spectral range included in our new study, the majority of the log$(gf)$-values available for comparison belong to \cite{ref:obrian1991}.

Our new work is the third in a series of articles published as a result of the collaboration between the Fourier Transform Spectroscopy laboratory at Imperial College London (IC) and the University of Wisconsin-Madison (UW). It completes the previous works on log$(gf)$ for Fe~I lines of interest in the Gaia-ESO survey \citep{ref:ruffoni2014} and oscillator strengths for transitions coming from high-lying even-parity Fe~I levels \citep{ref:denhartog2014}. In this paper we focus on the log$(gf)$ values for transitions coming from high-lying odd-parity upper energy levels, four of which contain spectral lines of particular interest for the Gaia-ESO survey. We provide new radiative lifetimes for 60 high-lying odd-parity levels obtained at the UW, 39 of which are measured for the first time.  Fe I emission spectra were recorded with the Fourier transform spectrometers at IC and at the National Institute of Standards and Technology (NIST). Measurements of branching fractions were completed for 15 of the previously mentioned odd-parity levels, and were combined with the new lifetimes to obtain 120 accurate values of oscillator strengths (and transition probabilities). Comparison with previous experiments shows that 22 of the analysed transitions have no earlier log$(gf)$-values and for 63 transitions the accuracy of the log$(gf)$s are improved compared with existing measurements in the literature.

\section{Experimental procedure}

Oscillator strengths, or absorption f-values, are obtained experimentally from
the measurement of atomic transition probabilities, $A_{ul}$, where
the subscript \textit{ul} refers to the transition from a given
upper energy level, \textit{u}, to a lower level, \textit{l}. Transition probabilities and absorption f-values are related by \citep{ref:spectrophysics}:

\begin{equation}
\label{E1}\log(g_lf) = \log\Bigl[A_{ul}g_u \lambda^2\times 1.499\times 10^{-14}\Bigl]
\end{equation}

\noindent where $g_l$ and $g_u$ are the statistical weights of the lower and upper energy level, respectively, and  $\lambda$ is the wavelength of the line expressed in nm.

The transition probability of a given atomic transition can be obtained spectroscopically,
since in the case of an optically thin plasma it is proportional to the area under the profile of the corresponding spectral line.
The integrated area of each intensity calibrated spectral line, $I_{ul}$, is proportional to its intensity in photons per second \citep{ref:pickering2001a}.

So called Branching Fractions \citep{ref:huber1986} are given by:

\begin{equation}
\label{eqn:bf} BF_{ul} = \frac{I_{ul}}{\sum_l I_{ul}}= \frac{A_{ul}}{\sum_l A_{ul}}
\end{equation}

\noindent and for a particular upper energy level the lifetime $\tau_u$ is:

\begin{equation}
\label{eqn:lifetime} \tau_u = \frac{1}{\sum_l A_{ul}}.
\end{equation}

\noindent As long as the sum of the $A_{ul}$ includes all branches to the lower energy
levels, we can combine expressions \ref{eqn:bf} and \ref{eqn:lifetime} to
obtain the transition probability of a given transition as:

\begin{equation}
\label{eqn:aki} A_{ul} = \frac{\mbox{BF}_{ul}}{\tau_u}
\end{equation}

\noindent As can be seen, this method of measuring $A_{ul}$ has the advantage
that no assumption needs to be made regarding the thermodynamic
equilibrium of the plasma used as a light source.

\subsection{Branching fraction measurements}

\begin{figure*}
\centering
\includegraphics[width=0.45\textwidth,trim=2.8cm 0 2.2cm 2cm]{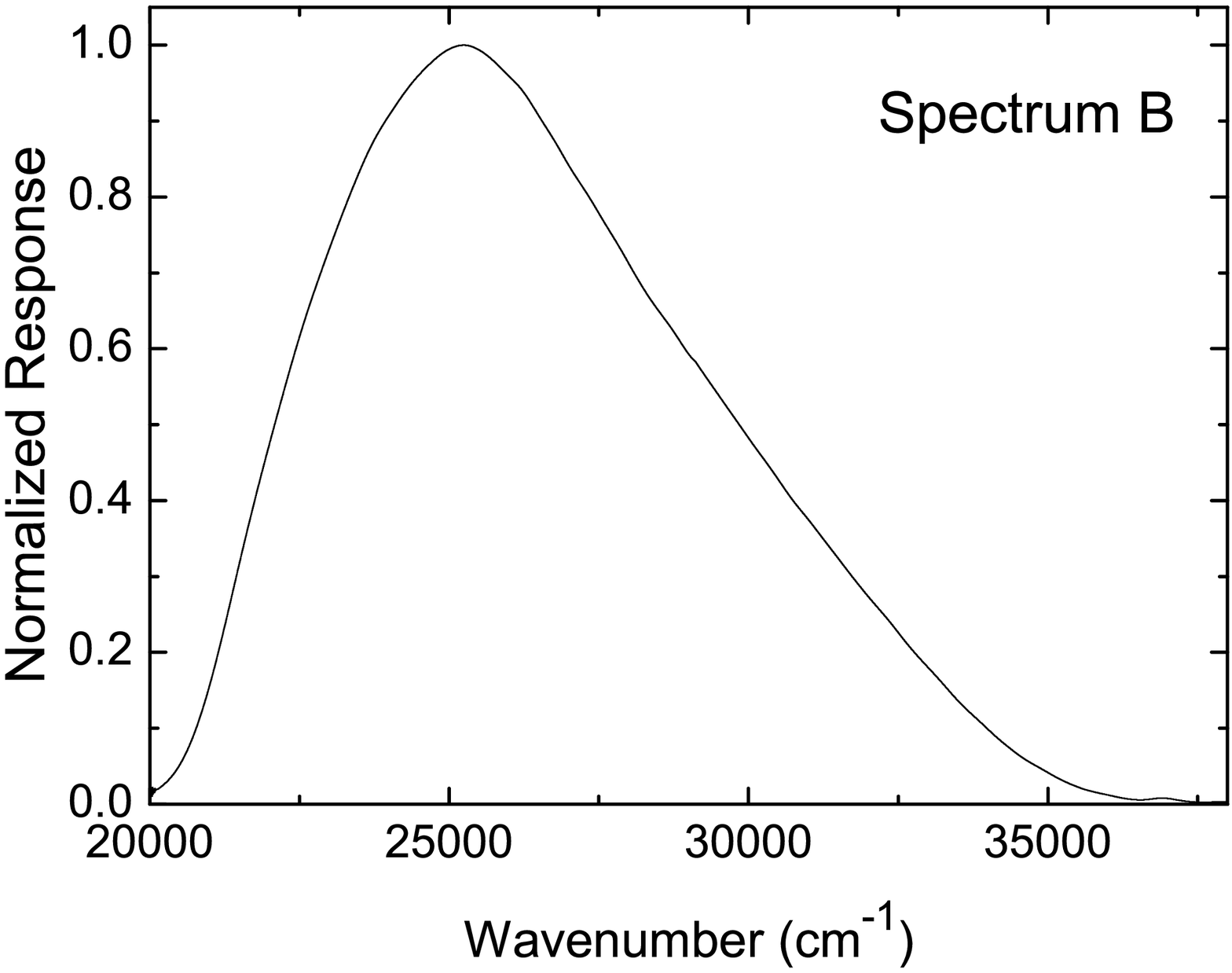}\includegraphics[width=0.45\textwidth,trim=2.8cm 0 2.2cm 2cm]{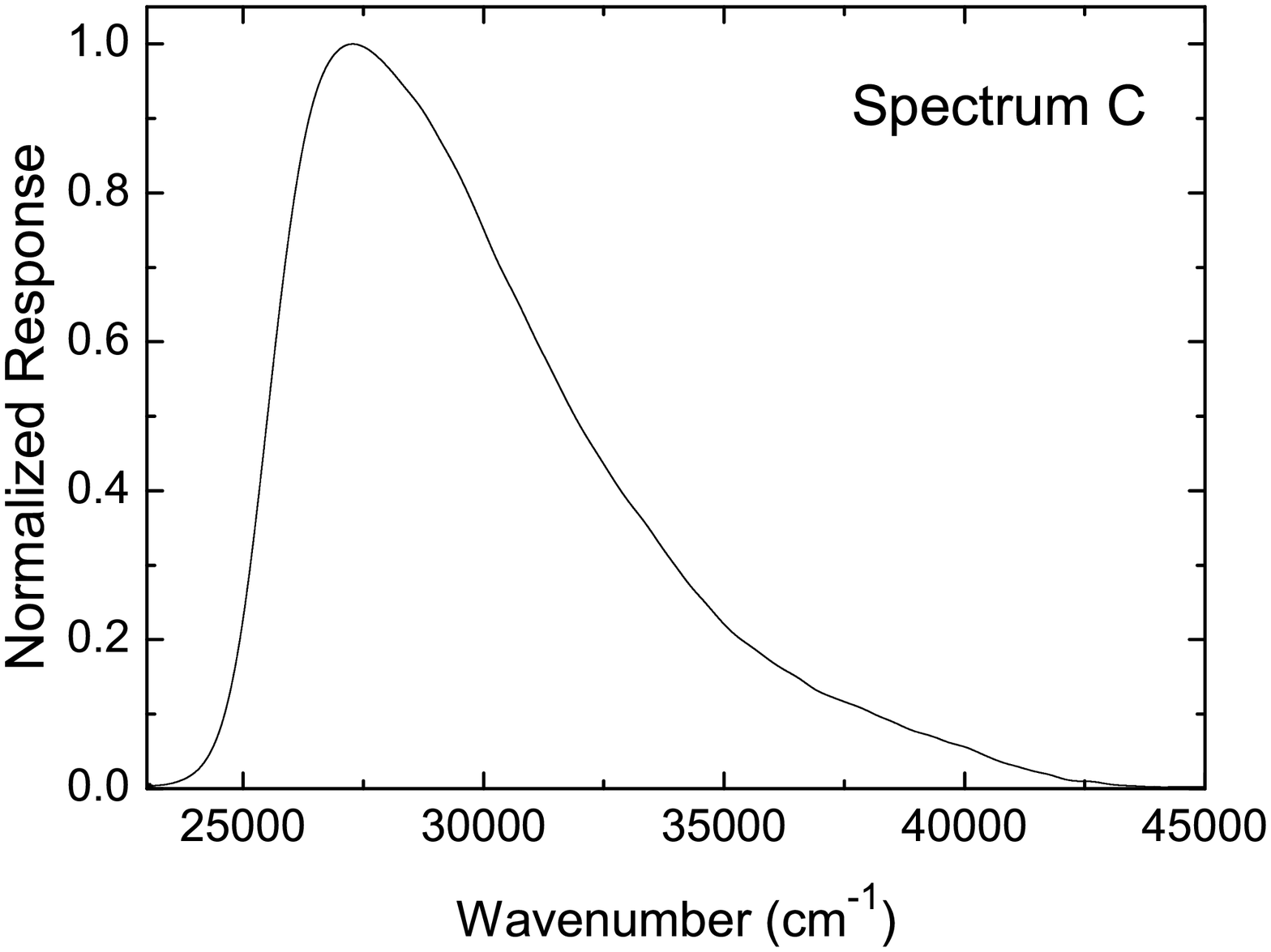}\\
\includegraphics[width=0.45\textwidth,trim=2.8cm 0 2.2cm 2cm]{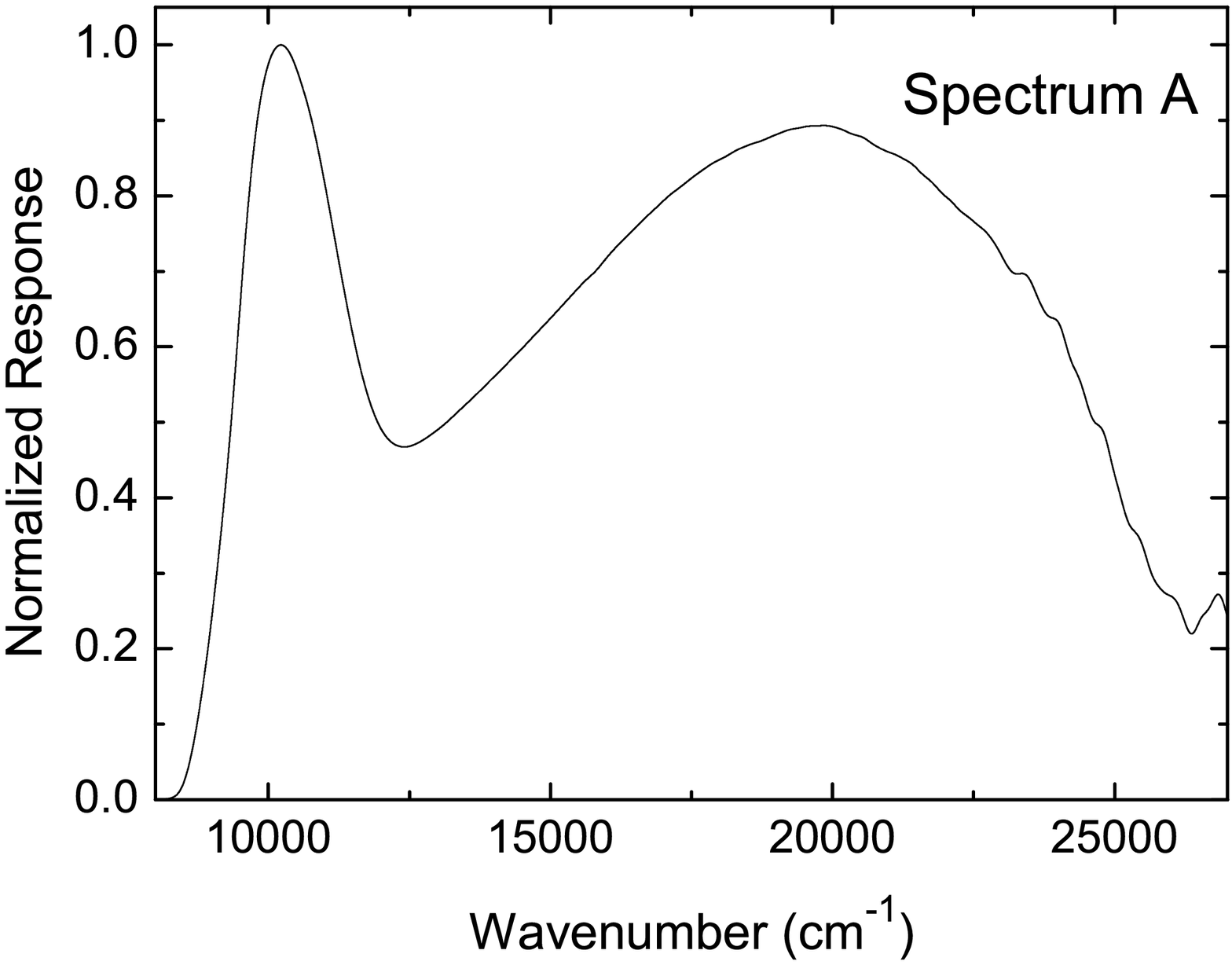}\includegraphics[width=0.45\textwidth,trim=2.8cm 0 2.2cm 2cm]{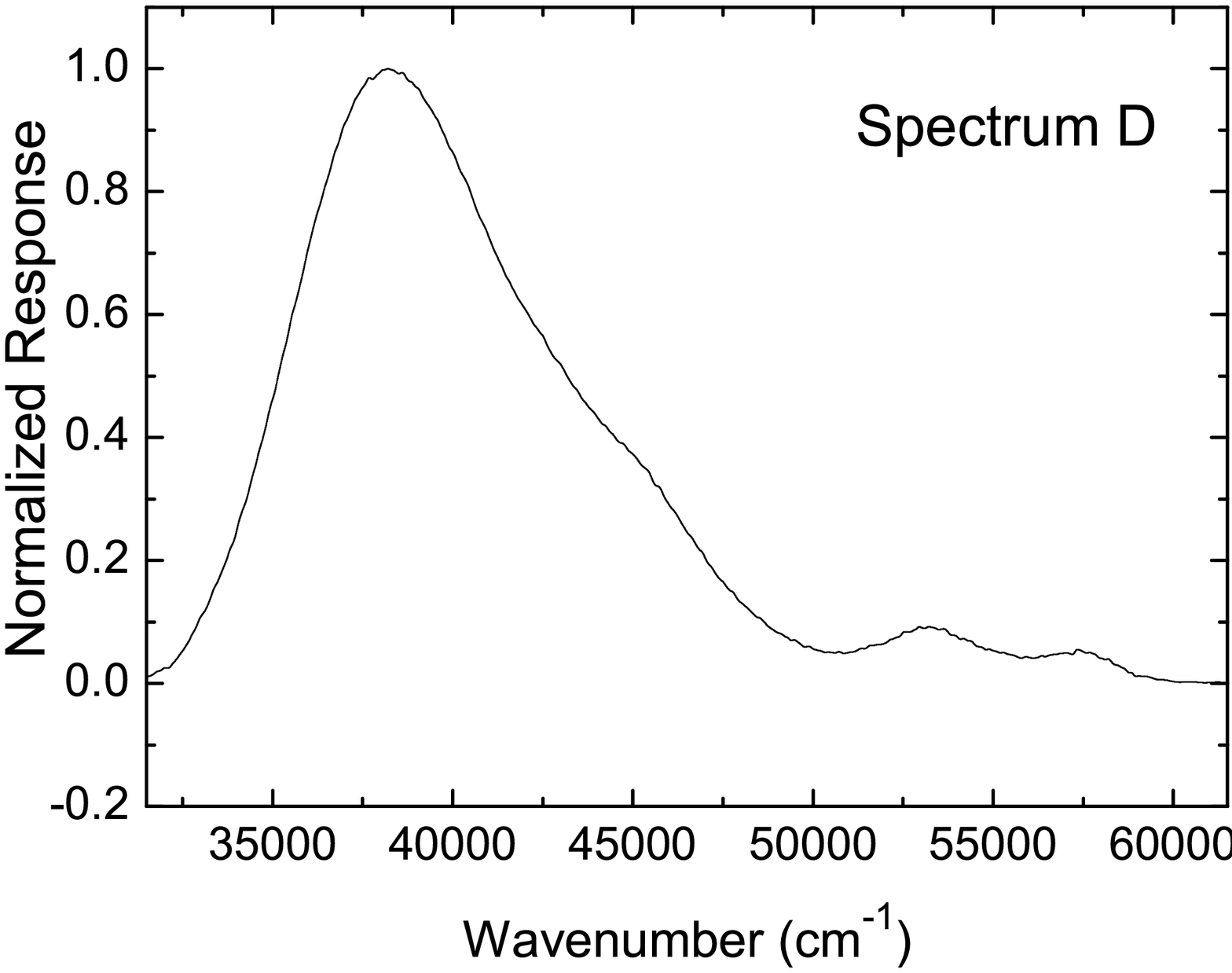}\\
\includegraphics[width=0.45\textwidth,trim=2.8cm 0 2.2cm 2cm]{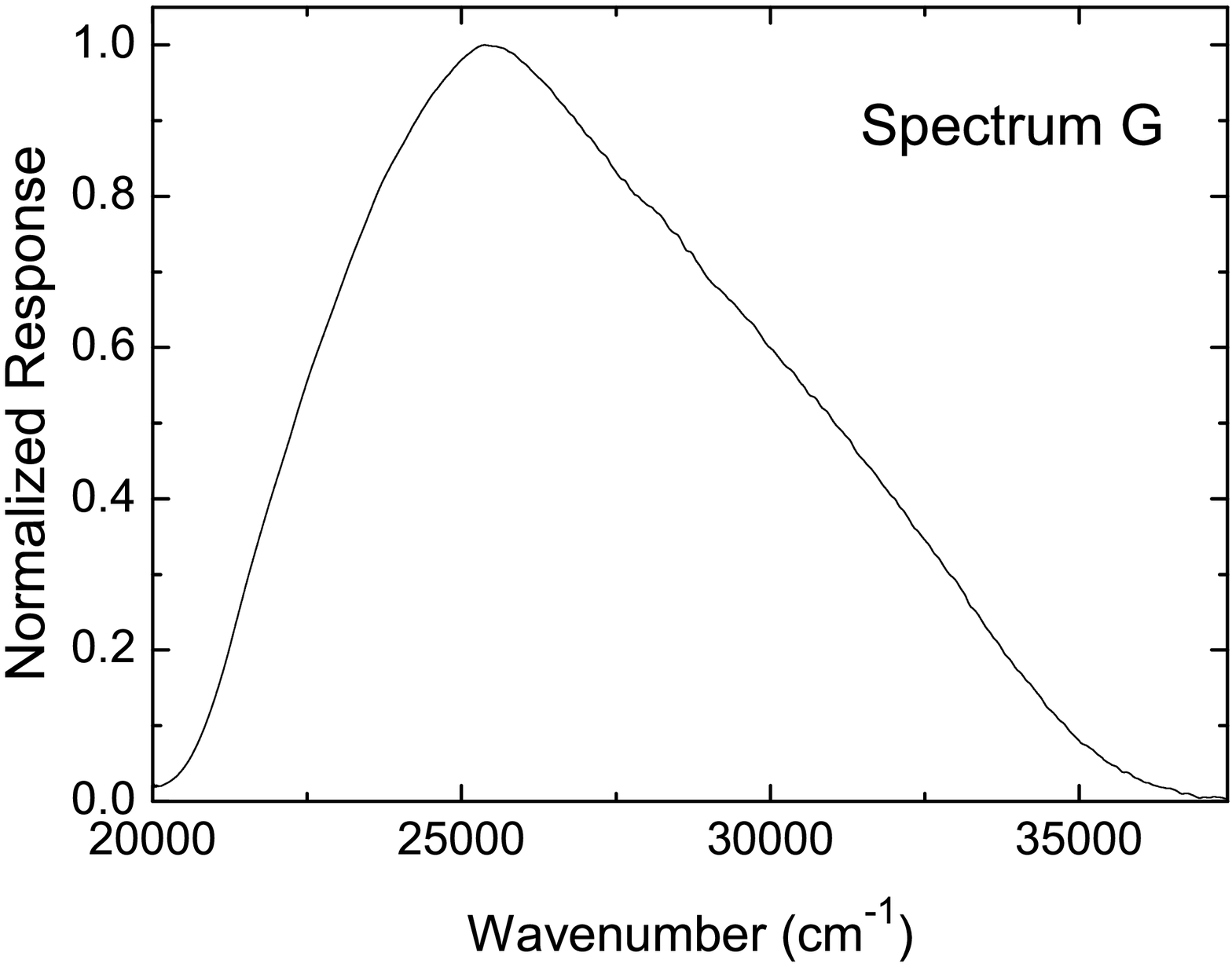}
\caption{Response functions used to intensity calibrate the five different spectra used in this work (see Table \ref{table:spectra}).}
\label{fig:response}
\end{figure*}

Two different sets of spectra were used to obtain the log$(gf)$ values included in this work.
Spectrum A was measured on the  2 m FT spectrometer at the National Institute of Standards and Technology (NIST) and it covers the
spectral range between 8000 and 26000 cm$^{-1}$. An iron cathode
mounted in a water cooled hollow cathode lamp (HCL) was used to
generate the plasma used as light source. The HCL was run in Ne at a
pressure of 2.1 mbar and a current 2 A. A detailed description of
this measurement can be found in \cite{ref:ruffoni2014} and
\cite{ref:denhartog2014}. The response function of the spectrometer, shown in Fig.\ref{fig:response} for Spectrum A, was obtained by using a calibrated Standard tungsten (W) halogen lamp in the spectral range between 250 and 2400 nm. In order to verify that this spectrometer response was stable over time, spectra of this tungsten lamp (whose radiance is known to $\pm$1.1\%) were measured before and after acquiring iron spectra with the HCL.

Spectra B, C, D and G were measured on the IC VUV Fourier Transform
Spectrometer (FTS) covering the spectral range between 20000 and 62000~cm$^{-1}$. The resolution, detector and filter used to record
each spectrum , as well as the experimental conditions are included in
Table~\ref{table:spectra}. The Fe I emission spectra were produced
in a water cooled HCL filled with Ne at a
pressure ranging from 1.3 to 1.4~mbar and with a  99.8\% pure iron cathode
operated as the source.  Currents of 700 or 1000~mA
(see Table~\ref{table:spectra}) were selected depending on the
signal-to-noise ratio of the spectral lines of interest. These conditions were
optimized to obtain the highest signal-to-noise ratio for the weaker
lines, whilst avoiding self-absorption effects for the stronger lines.

Two Standard intensity calibrated lamps were used to obtain the response function of
the IC VUV spectrometer for these four spectra: a deuterium lamp (D$_{2}$)
in the spectral interval ranging from 200 to 350 nm and a tungsten lamp (W)
for longwards of 300 nm. Spectra from these two lamps were measured before
and after each of the runs recorded with the HCL lamp and using the
same measurement conditions.
The uncertainties of the relative spectral radiance
of the W lamp, calibrated by the UK National Physical Laboratory (NPL) were lower than $\pm$1.4\% between 410 and 800 nm, increasing to $\pm$2.8\% at 300 nm. The deuterium lamp, calibrated by the Physikalisch-Technische Bundesanstalt (PTB) in Germany, has a relative spectral radiance with an uncertainty of $\pm$7\% between 170 and 410 nm.
The response functions of the different spectra, shown in Fig.\ref{fig:response}, were obtained by combining the response functions found using both lamps, W and D$_{2}$.

The Fe spectra were fitted by using Voigt profiles with the XGREMLIN
package \citep{ref:xgremlin}. Once fitted and intensity calibrated, the
spectra were used to obtain the branching fractions of all the
 transitions coming from the upper energy levels of
interest by using the FAST software package \citep{ref:ruffoni2013a}. The calculated branching fractions of \cite{ref:kurucz2007} were used to define the transitions to be included, and to estimate the completeness of the set of transitions from each upper level considered, as described in \cite{ref:pickering2001a} and \cite{ref:pickering2001b}.
 Final
results for new log$(gf)$s were also obtained using the FAST software to combine the branching fractions with the
new experimental upper energy lifetime values described in Section
2.2.

All the fitted spectral lines were checked for possible self-absorption by
carefully examining the residuals from the line fits. Possible cases of
blended lines were also analysed by checking the Fe~I linelist of \cite{ref:nave1994} and Fe~II linelist of \cite{ref:NaveJoh2013}, as well as the theoretical
log$(gf)$s of \cite{ref:kurucz2007}.

For most of the cases, more than one spectrum from
Table~\ref{table:spectra} was necessary to encompass all the
transitions coming from a given upper energy level. In these
situations, all the lines were put on a common relative intensity scale by
calculating the ratio in the intensity of several lines from that
particular upper energy level which were measured in the overlapping
region of the pairs of spectra. A detailed description of this
method can be found in \cite{ref:pickering2001a} and
\cite{ref:pickering2001b}.

We have paid special attention to the calculation of the
experimental uncertainties introduced in our BF measurements by
taking into account all the different sources of error, such as the
uncertainty in the signal-to-noise ratio of the observed Fe I spectral line and the
standard calibration lamp spectra, as well as the spectral radiance uncertainty of these standard light sources. A detailed description can be found in \cite{ref:ruffoni2013a} and
\cite{ref:ruffoni2013b}, but the general expression is included here
for completeness. The experimental uncertainty of a given BF can be defined as:

\begin{equation}
\Biggl(\frac{\Delta\mbox{BF}_{ul}}{\mbox{BF}_{ul}}\Biggr)^2 = (1 -
2\mbox{BF}_{ul})\Biggl(\frac{\Delta I_{ul}}{I_{ul}}\Biggr)^2 +
\sum_{j=1}^n\mbox{BF}_{uj}^2\Biggl(\frac{\Delta
I_{uj}}{I_{uj}}\Biggr)^2
\end{equation}

\noindent where $I_{ul}$ is the calibrated relative intensity of the emission
line associated with the electronic transition from level $u$ to
level $l$, and $\Delta I_{ul}$ is the uncertainty in intensity of
this line due to its measured signal-to-noise ratio and the
uncertainty in the intensity of the standard lamp. From Equation
\ref{eqn:aki}, it then follows that the uncertainty in $A_{ul}$
is:

\begin{equation}
\Biggl(\frac{\Delta A_{ul}}{A_{ul}}\Biggr)^2 = \Biggl(\frac{\Delta
\mbox{BF}_{ul}}{\mbox{BF}_{ul}}\Biggr)^2 + \Biggl(\frac{\Delta
\tau_{ul}}{\tau_{ul}}\Biggr)^2~
\end{equation}

\noindent where $\Delta \tau_{ul}$ is the uncertainty in our measured upper
level lifetime. Finally, the uncertainty in $\log(gf)$ of a given
line can be calculated as:

\begin{equation}
\Delta \log(gf) = \log\Biggl(1 +  \frac{\Delta
A_{ul}}{A_{ul}}\Biggr)~
\end{equation}

\subsection{Radiative lifetime measurements}

Radiative lifetimes provide the absolute scale for the branching
fractions.  They are measured using time-resolved laser-induced
fluorescence (TRLIF) on a slow beam of iron atoms. This beam is
produced from a hollow cathode discharge sputter source. A pulsed
electrical discharge is operated in $\sim$ 50~Pa argon gas at 30~Hz
repetition rate. The pulses have $\sim$10~A peak current and 10 $\mu$s
duration during which the energetic argon ions sputter atoms from
the pure iron foil lining the stainless steel hollow cathode. The
discharge is maintained between pulses with a 30~mA DC current. The
cathode is closed on the downstream end except for a 1 mm flared
hole, through which the gas phase iron is differentially pumped into
the low pressure ($\sim$10$^{-2}$~Pa) scattering chamber. The beam
is weakly-collimated and slow (the neutrals have speeds of $\sim$
5$\times$10$^{4}$~cm$.$s$^{-1}$ and the ions are somewhat faster), and contains both
neutral and singly-ionized iron in their ground and low metastable
levels.

The atomic/ionic beam is intersected at 90$^{\circ}$ angles by a
beam from a nitrogen laser-pumped dye laser. This intersection takes
place 1~cm below the bottom of the cathode. The laser pulse is
delayed relative to the peak of the discharge pulse by $\sim$
20~$\mu$s to account for the transit time of the atoms. The duration
of the laser is $\sim$ 3~ns (FWHM) and the pulse terminates
completely within a few nanoseconds. This latter characteristic
allows the fluorescence to be recorded free from laser interaction,
making it unnecessary to deconvolute the laser pulse and
fluorescence signals. The wavelength of the laser is tunable over
the range 205 - 720~nm using a large selection of dyes as well as
frequency doubling crystals. The narrow bandwidth of the laser
(0.2~cm$^{-1}$) ensures selective excitation of the level of
interest. Cascade radiation from higher-lying levels, which
troubled earlier, non-selective techniques such as beam foil
excitation, is not a problem in this experiment.

A transition is chosen for laser excitation which is classified to
the level of interest and has some observed intensity in the NIST
line list for Fe I \citep{ref:kramida2011}$^1$. The transition must
also originate in the ground or low-lying metastable levels which
are populated in the atomic beam. We find that neutral iron
metastable levels up to 25~000~cm$^{-1}$ have sufficient population
for use as lower levels for laser excitation. Care must be taken to
correctly identify the transition in the experiment, particularly
when working in such dense spectra as Fe I, II. We do not rely
on an absolute measurement of the laser wavelength. Rather, the
wavelength is course-tuned to within 0.1~nm of the transition by
adjusting the grating of the dye laser while monitoring the
wavelength with a 0.5 m focal length monochromator. An LIF spectrum
is then recorded while the laser wavelength is slowly changed over a
range of 0.5 - 1~nm. This fine control of the laser is accomplished
by pressurizing an aluminum box which houses the laser grating up to
1300~kPa of nitrogen and then slowly bleeding the nitrogen away,
changing the index of refraction. This low-tech method yields
extremely linear and reproducible control of the laser wavelength.
The separation between lines on the LIF spectrum can be measured
accurately to $\pm$0.002~nm. This spectrum is then pattern-matched
to the NIST line list to correctly identify the transition of
interest.

Once the laser wavelength is tuned to the transition, fluorescence
is collected at right angles to both laser and atomic beams through
a pair of fused-silica lenses. These lenses comprise an f/1 optical
system.  Allowance is made for the insertion of optical filters
between the two lenses where the fluorescence is roughly collimated.
These filters can be broadband colored-glass filters or narrowband
multi-layer dielectric interference filters. Their function is to
block scattered laser light, light from the discharge and cascade
radiation.  Although cascade from higher levels is not a problem due
to the selective nature of the excitation, cascade from lower-lying
levels is still a possibility. Fluorescence from the beam
interaction region is imaged onto the photocathode of a RCA 1P28A
photomultiplier tube (PMT). The PMT signal is then recorded with a
Tektronix SCD1000 transient digitizer beginning at least 7~ns after
the peak of the laser pulse. This delay allows time for the
complete termination of the laser so that deconvolution of the laser
temporal profile from the fluorescence signal is not necessary. An
average of 640 fluorescence decays is recorded. The laser
wavelength is then tuned off the transition and an average of 640
background traces is recorded. These data are downloaded to a
computer for analysis. The digitized data is divided into an
early-time and a late-time section for analysis, each being $\sim$
1.5 lifetimes in length. A least-squares-fit to a single
exponential is performed on the background subtracted signal to
determine a lifetime in each section. Comparison of the early- and
late-time lifetimes gives a quick and sensitive method to check for
any systematic deviations from a clean exponential. Five such
lifetime measurements are averaged together for a given set of
experimental conditions. Two measurements of each lifetime are made
with typically several months intervening and using a different
laser transition whenever possible. This redundancy ensures that
the experiment is running reproducibly, that the transitions are
identified correctly in the experiment, that they are classified
correctly to the upper level and that they are not masked by a
hidden blend or affected by cascade radiation through lower levels.

In addition to cascade radiation, there are several other effects
which must be understood and controlled to ensure a clean lifetime
measurement. The dynamic range of the experiment extends from
$\sim$ 2~ns to several microseconds. The bandwidth of the PMT,
digitizer and associated electronics begins to affect the fidelity
of the lifetime measurements below $\sim$ 4~ns and limits the
minimum lifetime to $\sim$ 2~ns. We assign a minimum uncertainty of
0.2~ns, such that the fractional uncertainty rises from 5\% at 4~ns
up to 10\% at 2~ns lifetime. The other end of the dynamic range is
limited by the flight-out-of-view effect, where the motion of the
atoms has taken those radiating later in the decay outside the view
of the PMT. This has the effect of artificially shortening the
measured lifetime. This effect can be mitigated somewhat by
inserting a cylindrical lens in the optical train which serves to
defocus the optics in the direction of motion, making them much less
sensitive to that motion. This step is taken for neutral lifetimes
$>$ 300~ns and ion lifetimes $>$ 100~ns (ions move somewhat faster
than the neutrals). It also has the unfortunate effect of
diminishing the signal levels by a factor of five or so. Zeeman
quantum beats arise when the atomic dipoles excited by the polarized
laser have time to precess in the earth’s magnetic field before
they radiate. To avoid this effect, the region where the laser and
atomic beams interact is placed at the center of a set of Helmholtz
coils which are used to zero the field to within $\pm$2~$\mu$T. This
tolerance is adequate to avoid Zeeman quantum beats for shorter
lifetimes, but for longer lifetimes ($>$300 ns) some effect can
still be observed. In these cases a high magnetic field (3~mT) is
produced with a second set of coils which causes rapid precession
and the Zeeman beats are washed out on the longer digitizer time
windows employed. A further systematic effect arises from
after-pulsing in the PMT. Generally, the characteristics of the
1P28A PMT, i.e. fast rise-time and high sensitivity in the UV and
visible, are favorable for lifetime measurements. However, the PMT
does produce a weak (0.1\%) after-pulse as a result of the prompt
electron cascade ionizing residual gas in the tube. This weak and
relatively slow signal is picked up on the photocathode and results
in a systematic, reproducible lengthening of lifetimes around
100~ns. This effect of a few percent is corrected for in the final
lifetimes.

We periodically measure a set of benchmark lifetimes which helps us
ensure that the experiment is running reproducibly and accurately. These benchmarks are lifetimes which are well known from other sources. Some are
from theoretical calculations and others from experiments which have
smaller, and generally different systematic uncertainties than our
own. The benchmarks measured for the current set of lifetimes are:
z$^6$F$_{11/2}$ and z$^6$D$_{9/2}$  states of Fe$^+$ at 3.19(4)~ns
and 3.70(6)~ns, respectively (laser-fast beam,
\cite{ref:biemont1991}), 2$^2$P$_{3/2}$ state of Be$^+$ at
8.8519(8)~ns (variational method calculation,\cite{ref:yan1998});
the 3$^2$P$_{3/2}$ state of neutral Na at 16.23(1) ns (NIST
critical compilation of \cite{ref:kelleher2008}); 4p'[1/2]$_1$ state
of Ar at 27.85(7)~ns (beam-gas-laser-spectroscopy,
\cite{ref:volz1998}); 3$^3$P, 4$^3$P, and 5$^3$P states of neutral
He at 94.8(1) and 219.3(2)~ns (variational method calculation,
\cite{ref:kono1984}; \cite{ref:drake2007}). These benchmarks allow
us to quantify small corrections due to residual systematic effects,
ensuring that our lifetimes are well within the stated uncertainty.
A comparison of our lifetimes to laser-fast beam measurements
performed by Scholl et al. (2002) in Sm II suggests that the stated
uncertainties are conservative \citep{ref:lawler2008}.

\section{Results and discussion}

In Table \ref{table:lifetimes} we report the results of lifetime
measurements from this study as well as LIF results from the
literature.  Lifetimes measured using older, less reliable
techniques are not listed. Lifetimes are given for 60 high-lying
odd-parity levels of Fe I ranging in energy from 27~166.82 to 57~565.31~cm$^{-1}$.
The fractional uncertainty in our measurements is
5\%, except for those lifetimes less than 4~ns for which the
absolute uncertainty is 0.2~ns. Approximately two-thirds of these
lifetimes are measured for the first time. The comparison with the
earlier work is very favorable. The lifetimes
in the \cite{ref:obrian1991} work were measured in our (UW) lab
with nearly the same apparatus as the current work (in the earlier
work a different digitizer was used). We re-measured some of the
original \cite{ref:obrian1991} lifetimes to ensure that, even after
25 years, the experiment is giving consistent results.  Happily, we
see very good agreement with this older work. For eight lifetimes
in common, the mean and rms differences between the current study
and \cite{ref:obrian1991} are -1.5\% and 3.1\%, respectively, using
the current study as reference (in the sense (theirs $–-$
ours)/ours). The level of agreement with \cite{ref:engelke1993} is
also very good with mean and rms differences of -1.1\% and 5.4\%,
respectively. Our study overlaps with that of \cite{ref:marek1979}
for only two lifetimes which agree within 0.5\% for the longer
lifetime around 64 ns and within 4.5\% for a shorter lifetime
around 9~ns. Our study overlaps with that of \cite{ref:langhans1995} for only two very short lifetimes less than 3~ns. We agree perfectly in one case and differ by only 0.1~ns for the other. The excellent level of agreement with these four earlier studies is typical of modern TRLIF measurements.

Measurements of branching fractions were attempted for all the transitions coming from the upper energy levels included in Table~\ref{table:lifetimes} and completed for 15 of them. These upper energy levels are listed in Table~\ref{table:completeness} together with their configuration, the lifetime of the level used for the determination of the log$(gf)$s and the completeness of each set of transitions. The remaining energy levels were excluded from our study due to the impossibility of putting the different lines onto a common intensity scale or to the presence of blended or very low signal-to-noise ratio lines. Table~\ref{table:results} lists the results for branching fractions,
transition probabilities and log$(gf)$s with their uncertainties for 120 transitions of Fe I, 22 of which are new and 63 have improved uncertainties compared with existing data. The spectral lines are
grouped by common upper energy level and each set is sorted in
order of descending wavelength. The values of the air wavelengths,
as well as the upper and lower energy levels and J were taken from
\cite{ref:kramida2011} based on \cite{ref:nave1994}. In addition, the log$(gf)$s measured in this experiment are compared when possible with the value recommended by \cite{ref:fuhr2006},
included in the last column along with the reference from the work
from which the oscillator strength was taken. The letters L and P are used to indicate the method used in the O'Brian work to obtain these values and stand for `lifetime' and `population method', respectively.

\begin{figure*}\centering
\includegraphics[width=150mm, trim=3cm 0 3cm 2cm]{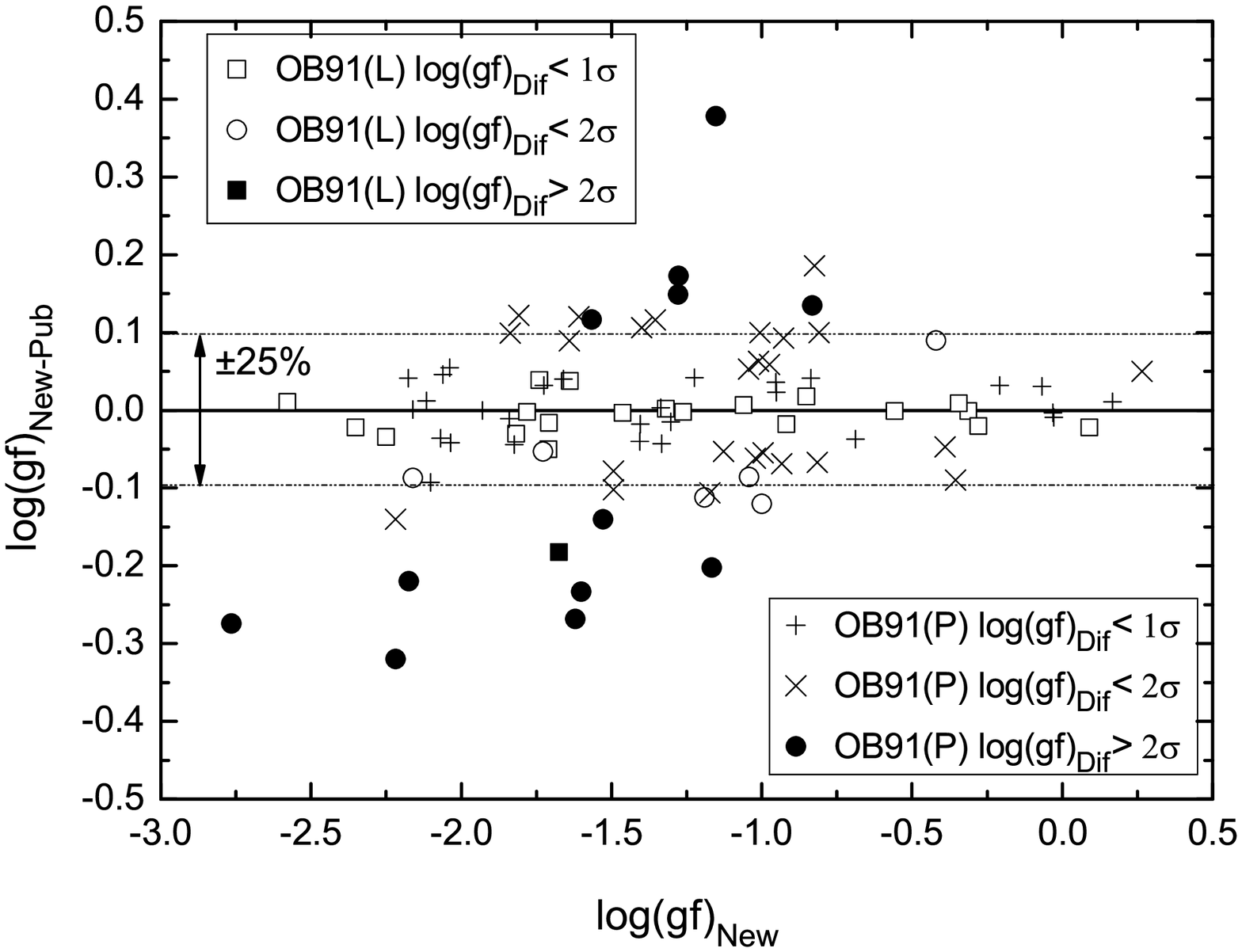}
\caption{Comparison of the new log$(gf)$ values measured in this work with those of \cite{ref:obrian1991} obtained from lifetimes (L) and extrapolating energy level populations (P). The solid horizontal line represents perfect agreement between the two sets of values. The dashed horizontal line indicates uncertainties of $\pm$25\%, coded as `C' by \cite{ref:fuhr2006}. Agreement within the combined experimental uncertainty is indicated by $<1\sigma$.}
\label{F1}
\end{figure*}

\begin{figure*}\centering
\includegraphics[width=150mm, trim=3cm 0 3cm 2cm]{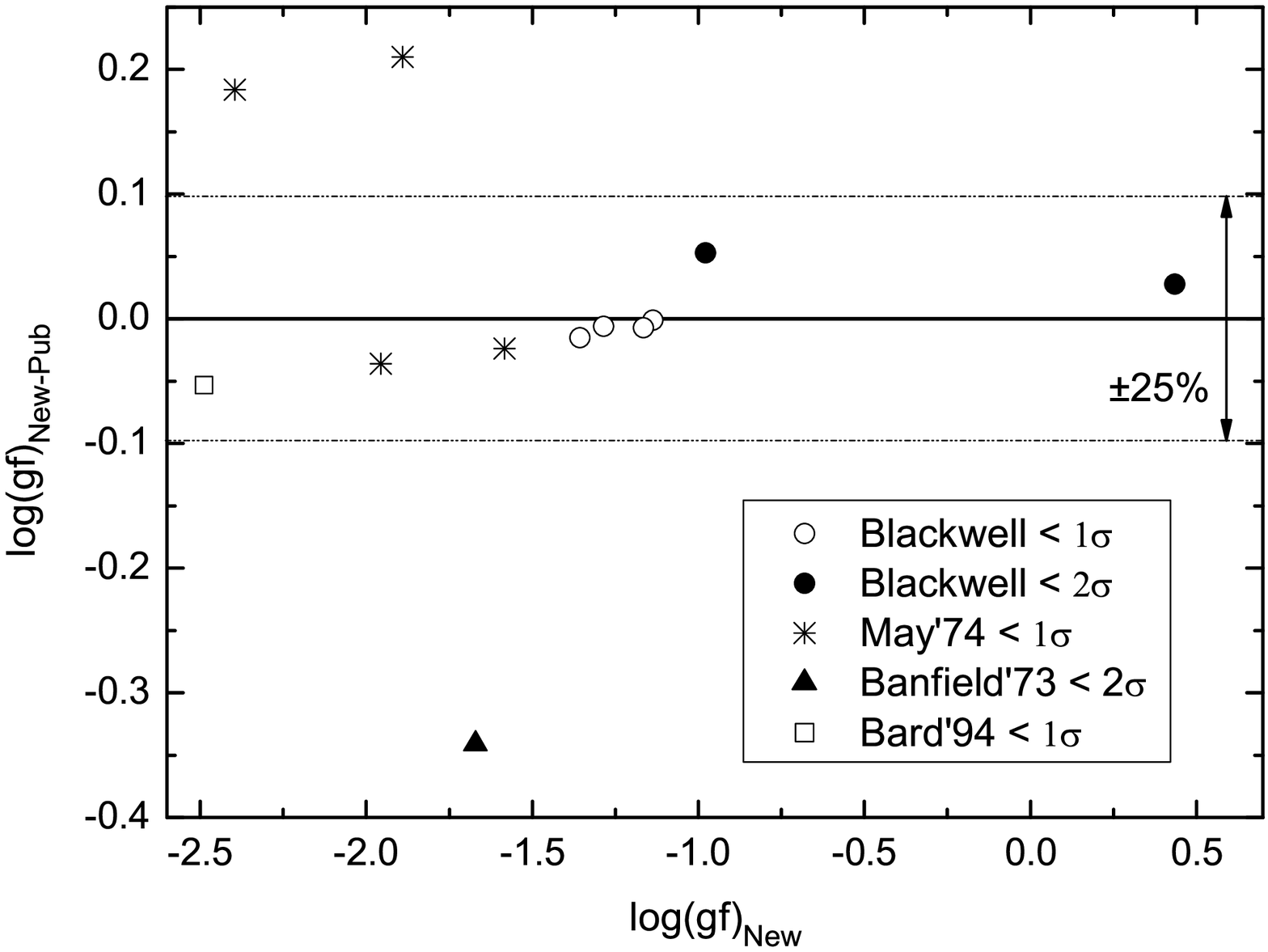}
\caption{Comparison between our new oscillator strengths and those from Blackwell et al. (1979, 1982a, 1982b), \cite{ref:may1974}, \cite{ref:bard1994} and \cite{ref:banfield1973}. Close agreement with Blackwell et al. is used as an indication of the quality of our new data.}
\label{F2}
\end{figure*}

Figs.~\ref{F1} and \ref{F2} show the comparison between the new log$(gf)$ values obtained in this experiment and those published previously, indicating if they agree within one or two $\sigma$, which represents the combined experimental uncertainty. Both plots include  dashed horizontal lines indicating uncertainties of $\pm$25\%, which correspond to values classified as `C' by \cite{ref:fuhr2006}, regarding their accuracy. In Fig.~\ref{F1}, our log$(gf)$s are compared to those from \cite{ref:obrian1991}. We have made a distinction between the log$(gf)$s which were obtained from measurements of lifetimes and branching fractions and those determined from extrapolated energy level populations and relative line intensities. For 13 of the compared transitions, marked in Fig.~\ref{F1} by filled symbols, \cite{ref:obrian1991} log$(gf)$s do not agree with our new values within 2$\sigma$. It is noted that only one of these values from \cite{ref:obrian1991} was obtained from lifetime measurements, with the remaining 12 being determined by using the population method.
\
Fig.~\ref{F2} shows the comparison with log$(gf)$ values from Blackwell et al. (1979, 1982a, 1982b), \cite{ref:may1974}, \cite{ref:bard1994} and \cite{ref:banfield1973}. It is possible to see how log$(gf)$s from Blackwell et al. agree within 1$\sigma$ with our new values, which is reassuring as their experiments are considered to be the most precise by \cite{ref:fuhr2006}, with uncertainties lower than 2\%. Relative log$(gf)$s from Blackwell et al. are claimed to be better than 2\%. The comparison with log$(gf)$s from \cite{ref:may1974}, obtained from emission measurements from a wall-stabilized arc, shows a wider scatter, which is not strange given the large uncertainties assigned by \cite{ref:fuhr2006}. The only log$(gf)$ value from \cite{ref:bard1994} available for comparison shows a good agreement within 1$\sigma$, whereas the log$(gf)$ from \cite{ref:banfield1973} differs significantly from our value, although the difference lies within 2$\sigma$.

\section{Solar spectral synthesis}

%
%
A subset of the new $gf$-values measured in this work were used to determine iron line abundances from the solar spectrum obtained with the Kitt Peak Fourier Transform Spectrometer by \cite{ref:kurucz1984}. 
The high-quality observations and the well-known atmospheric parameters available for the Sun make it an ideal test case to evaluate the impact of new atomic data on stellar spectral synthesis.
We selected 17 lines from Table~\ref{table:results} which
show low contamination from blends at the spectral resolution of the
solar flux atlas, and which are located in regions of good continuum
placement. The observed spectrum has a minimum wavelength of 300~nm,
and it is very crowded at wavelengths below $\sim$400~nm, which
excludes about two thirds of the lines published in this work. Among
the remaining lines, about half were too blended to attempt any
abundance derivation.
The selected lines are listed in Table~\ref{tab:solar} together with the required input atomic data in columns 1 to 4 (central wavelength, lower level energy $E_{\rm low}$, $gf$-value, and van der Waals parameter). The latter is used to account for line broadening due to collisions with neutral hydrogen and was extracted from the VALD database (\citealt{ref:kupka1999}, \citealt{ref:heiter2008}). 
The meaning of the values for the van der Waals broadening parameter given in Table~\ref{tab:solar} (column 3) is as follows. Values greater than zero were obtained from Anstee, Barklem \& O'Mara (ABO) theory (\citealt{ref:anstee1991}, \citealt{ref:anstee1995}, \citealt{ref:barklem2000}) and are expressed in a packed notation where the integer component is the broadening cross-section, $\sigma$, in atomic units, and the decimal component is the dimensionless velocity parameter, $\alpha$. Values less than zero are the log of the broadening parameter, $\gamma_6$ (rad s$^{-1}$), per unit perturber number density, $N$ (cm$^{-3}$), at 10\,000~K (i.e. log[$\gamma_6/N$] in units of rad s$^{-1}$ cm$^3$) from \cite{ref:kurucz2014}. 
These were used only when ABO data were unavailable. See \cite{ref:gray2005} 
for more details.

%
%
The spectral synthesis was done with the one-dimensional, plane-parallel radiative transfer code SME (\citealt{ref:valenti1996}, \citealt{ref:piskunov2017}, version 531) 
assuming local thermodynamic equilibrium (LTE) and using a model atmosphere interpolated in the MARCS grid included in the SME distribution \citep{ref:gustafsson2008}. 
We adopted an effective temperature of 5772~K and a logarithmic surface gravity (cm~s$^{-2}$) of 4.44 (\citealt{ref:heiter2015a}, \citealt{ref:prsa2016}), 
a microturbulence of 1.0~km~s$^{-1}$ and a projected rotational velocity of $v_{\rm rot}$sin($i$) = 1.6~km~s$^{-1}$ \citep{ref:valenti1996}. 
The instrumental profile was assumed to be Gaussian, with a width corresponding to the spectral resolution of the observations ($R$=200\,000).
Each line profile was fitted individually using $\chi^2$-minimization between observed and synthetic spectra while varying the iron abundance and the macroturbulence broadening, parameterised by a velocity $v_{\rm macro}$ in the radial-tangential model \citep{ref:gray2005}. 
In addition, a small radial velocity correction was applied to each line, allowing for variations in the wavelength scale of the observations.
For each line we determined the region of the line profile which seemed to be free from blends in the observed spectrum, and only spectrum points within that region were used to calculate the $\chi^2$.

%
%
The results are given in Table~\ref{tab:solar}, where we list the best-fit iron abundance log($\varepsilon$) for each line (column 10) on the standard astronomical scale\footnote{log($\varepsilon$) = log$_{10}(N_{\rm Fe}/N_{\rm H})$ + 12, where $N_{\rm Fe}$ and $N_{\rm H}$ are the number densities of iron and hydrogen atoms, respectively.}.
We also list the $v_{\rm macro}$ values derived for each line (column 9), as well as a measure for the goodness of fit (RMS deviation, column 11), ranging from 0.4 to 1.9 percent. The largest deviations are found for lines at wavelengths at or below 450~nm, and the smallest RMS values for lines above 700~nm.
Most of the abundances fall in the range from 7.3 to 7.6, with associated macroturbulence values between 2.5 and 3.7~km~s$^{-1}$ and radial velocity corrections of 0.2 to 0.4~km~s$^{-1}$.
However, the line at 4709~\AA has $v_{\rm macro} \gtrsim 4$ and an abundance of 7.9, significantly higher than the above range. This indicates contamination by undetected blends that have not been taken into account in the fit.
Excluding this line, the mean abundance and standard deviation are log$(\varepsilon)_{\rm new}$ = 7.47 $\pm$0.10~dex\footnote{The unit dex stands for decimal exponent, $x$ dex = $10^x$.} (16 lines), which is similar to the values found in the previous two papers in this series (7.44 $\pm$0.08~dex in \citealt{ref:ruffoni2014} 
and 7.45 $\pm$0.06~dex in \citealt{ref:denhartog2014}), 
and agrees with recent publications, such as 7.43 $\pm$0.05 from \cite{ref:bergemann2012} (MARCS LTE result), 
and 7.40 $\pm$0.04 from \cite{ref:scott2015} (mean of MARCS LTE abundances in their Table 1). 

The line-to-line abundance scatter might be influenced by non-LTE effects, which are however expected to be small in solar-like atmospheres. Unfortunately non-LTE corrections have only been published for few of the lines analysed here. Four of the lines were investigated by \cite{ref:gehren2001}, 
who derived non-LTE$-$LTE abundance differences of 0.03~dex for the 4495 and 5380~\AA\ lines, 0.05~dex for 4448~\AA, and $-0.11$~dex for 4433~\AA. Applying these corrections would lead to a slightly larger scatter. On the other hand, the more recent calculations by \cite{ref:bergemann2012} 
and \cite{ref:lind2012}, 
made available through the INSPECT database\footnote{http://www.inspect-stars.com} include two of the lines (4495, 5380~\AA), both with non-LTE corrections of 0.01~dex. If similar corrections apply to the remaining lines then they do not have any impact on the abundance scatter derived here.

\begin{figure*}\centering
\includegraphics{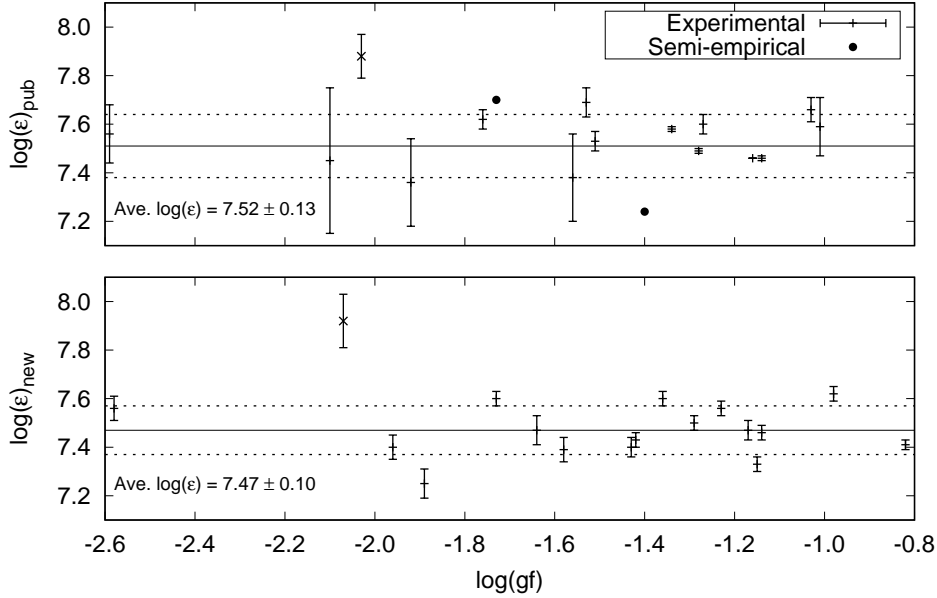}
\caption{Solar iron abundance, log($\varepsilon$), obtained from the synthesis of individual lines listed in Table~\ref{tab:solar} using the log($gf$) values from this work, log$(gf)_{\rm new}$ (lower panel), and the best previously published values, log$(gf)_{\rm pub}$ (upper panel). The point in each panel marked with a cross may be affected by undetected blends (see text) and was excluded from the calculation of the average log$(\varepsilon)$. The solid and dotted horizontal lines in each panel indicate the unweighted average abundance and the standard deviation, respectively. The error bars indicate only the uncertainty of the experimental $gf$-values (columns 5 and 7 in Table~\ref{tab:solar}) and do not capture the uncertainties associated with the solar atmospheric modeling and spectral synthesis.
\label{fig:solar}}
\end{figure*}

We repeated the abundance determination for the same set of lines using the best previously published experimental or theoretical log($gf$) values (see Table~\ref{tab:solar} for values and references, in columns 6 to 8, and for results in columns 12 and 13). The regions of the line profiles used for the fit were the same as above, and the macroturbulence values were those derived in the analysis with the new log($gf$) values\footnote{Simultaneous variation of macroturbulence resulted in the same $v_{\rm macro}$ values, except for two lines: 5533~\AA, and 7308~\AA, with 7\%, and 9\% lower values, respectively.}
The results for both the previously published data and the new data are illustrated in Fig.~\ref{fig:solar}.
The differences in derived abundances are consistent with the differences in log($gf$), and are larger than 0.05~dex for six lines\footnote{4548, 5380, 5533, 7220, 7308, 7443~\AA.}.
The mean abundance and standard deviation for previous data including the same 16 lines as above are log$(\varepsilon)_{\rm pub}$ = 7.52 $\pm$0.13~dex, which is close to log$(\varepsilon)_{\rm new}$, although slightly offset towards higher abundances and with a somewhat larger scatter.
In summary, the small scatter in the line abundances derived with the new data, and the satisfactory agreement with recently published values for the solar iron abundance validates the general accuracy of the new measurements.

\bibliography{solar_synthesis}
%

%
%

\section{Conclusions}

We report radiative lifetimes for 60 odd-parity energy levels ranging from 27~166 to 57~562~cm$^{-1}$ , 39 of which had not been measured before.  The uncertainties for these lifetimes are the larger of $\pm$5\% or 0.2~ns. When values are available in the literature, our results are in good agreement with them.

We provide 120 experimental log$(gf)$ values (transition probabilities) for Fe I transitions within the spectral range 213-1033 nm coming from 16 upper energy levels, 24 of which had no previous experimental data.  The uncertainty of these oscillator strengths has been carefully calculated and it ranges between 0.02~dex for the strongest lines and 0.09~dex for the very weak ones.
This accuracy is an improvement over previous log$(gf)$ measurements for around 72 of the transitions, making our new log$(gf)$ values good candidates for use in the analysis of stellar spectra and the determination of chemical abundances. Our log$(gf)$ results are in good agreement with those of Blackwell et al. (1979, 1982a, 1982b) and in general with \cite{ref:obrian1991} data.

\section*{Acknowledgements}

MTB, JCP and MPR would like to thank the UK Science and Technology Facilities Council (STFC) for support under research grant ST/N000838/1. EADH, JEL and AG would like to acknowledge funding from the National Science Foundation through award AST-1211055 and through the NSF REU program award AST-0907732. U.H. acknowledges support from the Swedish National Space Board (SNSB/Rymdstyrelsen). This work has made use of the VALD database, operated at Uppsala University, the Institute of Astronomy RAS in Moscow, and the University of Vienna. This work has made use of the NSO/Kitt Peak FTS data, produced by NSF/NOAO.


\begin{deluxetable}{lccccccl}
\tablewidth{0pt}
\tabletypesize{\scriptsize}
\tablecaption{FTS spectra used for the branching fraction measurements.}

                                                                                                            \tablehead{
\colhead{Spectrum} & \colhead{Wavenumber} & \colhead{Resolution}    &\colhead{Detector}  &  \colhead{Filter}    &   \colhead{P gas} &   \colhead{I lamp}    &   \colhead{Spectrum}  \\
    &   \colhead{range used (cm$^{-1}$)} &   \colhead{(cm$^{-1}$)}   &       &       &       \colhead{(mbar)}    &       \colhead{(mA)}  &       \colhead{filename\tablenotemark{a}}
    }                                                                                                       \startdata
A (NIST)    &   $9600 - 26000$  &   $0.02$  &   Si diode    &   None    &   2.1 &   2000    &   Fe0301to0403\_Calib \\
B (IC)  &   $21000 - 33000$ &   0.037   &   R11568 PMT  &   Schott BG3  &   1.3 &   700 &   F130610.002.047\_Scaled \\
C (IC)  &   $23000 - 41000$     &   0.037   &   R11568 PMT  &   UG5 &   1.4 &   700 &   Fe130624.011.039\_Scaled    \\
D (IC)  &   $31000 - 47000$ &   0.037   &   R7154 PMT   &   None    &   1.3 &   1000    &   Fe130603.021.059\_Calib \\
G (IC)  &   $20000 - 35000$ &   0.037   &   R11568 PMT  &   Schott BG3  &   1.3 &   1000    & Fe130604A.007.034 
\enddata

\tablenotetext{a}{Several spectra were coadded to improve de signal-to-noise ratio of the spectral lines.}
\label{table:spectra}

\end{deluxetable}


\begin{deluxetable}{ccccccc}																							
\tablewidth{0pt}																							
\tabletypesize{\scriptsize}																							
\tablecaption{Radiative lifetimes of Fe I odd-parity levels including many higher levels studied for the first time in our work.  The uncertainty in our measurements is the larger of $\pm$5\% or $\pm$0.2 ns}		
															
\tablehead{																							
\colhead{Configuration$^{a}$}	&	\colhead{Term$^{a}$}		&	\colhead{J}				&		\colhead{Level$^{a}$}				&		\colhead{Laser Wavelengths$^{a}$}	&	\multicolumn{2}{c}{Lifetime (ns)}			\\
\cline{6-7}																						
&	&	& \colhead{(cm$^{-1}$)}	&\colhead{(nm)}	&	\colhead{This Expt.}				&		\colhead{Other Expt.}												
}																							
\startdata																							

3d$^6$($^5$D)4s4p($^3$P$^o$)						&	z $^5$F$^o$	&	4	&	27166.82	&	505.1634, 514.2928	&	63.4	&	63.6$^b$, 66.6$^c$, 63.7(4.0)$^d$		\\				
						&		&		&		&		&		&			\\				
3d$^6$($^5$D)4s4p($^3$P$^o$)						&	z $^5$P$^o$	&	3	&	29056.324	&	344.0605, 349.0573	&	43.7	&			\\				
3d$^6$($^5$D)4s4p($^3$P$^o$)						&	z $^5$P$^o$	&	2	&	29469.024	&	344.0988, 347.5449	&	43.2	&			\\				
3d$^6$($^5$D)4s4p($^3$P$^o$)						&	z $^5$P$^o$	&	1	&	29732.736	&	344.3876, 347.6701	&	42.6	&			\\				
						&		&		&		&		&		&			\\				
3d$^7$($^4$F)4p						&	z $^5$G$^o$	&	6	&	34843.957	&	358.1192	&	9.2	&	9.6(0.6)$^d$		\\				
						&		&		&		&		&		&			\\				
3d$^6$($^5$D)4s4p($^1$P$^o$)						&	x $^5$D$^o$	&	3	&	39969.853	&	250.1131, 445.9117	&	2.7	&	2.6$^b$, 2.8$^c$, 2.7(1)$^e$		\\				
3d$^6$($^5$D)4s4p($^1$P$^o$)						&	x $^5$D$^o$	&	1	&	40404.518	&	251.8101, 444.7716	&	2.8	&	2.6$^b$, 2.9$^c$, 2.7(1)$^e$		\\				
						&		&		&		&		&		&			\\				
3d$^5$($^6$S)4s$^2$4p						&	y $^7$P$^o$	&	4	&	40421.938	&	247.3156	&	317	&	309$^b$		\\				
						&		&		&		&		&		&			\\				
3d$^6$($^3$F$_2$)4s4p($^3$P$^o$)						&	v $^5$D$^o$	&	4	&	44022.525	&	227.0862, 229.2524	&	96.5	&	95.9$^b$		\\				
3d$^6$($^3$F$_2$)4s4p($^3$P$^o$)						&	v $^5$D$^o$	&	1	&	44760.746	&	226.9098, 228.3303	&	21.3	&	21.4$^b$		\\\\								
3d$^6$($^3$F$_2$)4s4p($^3$P$^o$)						&	w $^5$F$^o$	&	4	&	44415.074	&	225.0790, 227.2069	&	41.6	&	41.2$^b$		\\\\							
3d$^7$($^4$P)4p						&	y $^5$S$^o$	&	2	&	44511.812	&	226.7084, 273.6963	&	13.3	&	13.6$^b$		\\\\							
3d$^7$($^4$P)4p						&	w $^5$P$^o$	&	3	&	46137.097	&	216.6773, 218.6486	&	3.2	&			\\				
3d$^7$($^4$P)4p						&	w $^5$P$^o$	&	2	&	46313.537	&	219.1839, 220.0724	&	3.2	&			\\				
3d$^7$($^4$P)4p						&	w $^5$P$^o$	&	1	&	46410.381	&	218.7194, 219.6041	&	3.5	&			\\				
						&		&		&		&		&		&			\\				
3d$^6$($^3$P$_2$)4s4p($^3$P$^o$)						&	z $^3$S$^o$	&	1	&	46600.818	&	218.6892, 219.1204	&	16.5	&			\\				
						&		&		&		&		&		&			\\				
3d$^6$($^3$P$_2$)4s4p($^3$P$^o$)						&	y $^3$P$^o$	&	2	&	46727.074	&	215.8629, 344.7277	&	19	&			\\				
3d$^6$($^3$P$_2$)4s4p($^3$P$^o$)						&	y $^3$P$^o$	&	1	&	46901.832	&	217.2584, 217.6840	&	11	&			\\				
						&		&		&		&		&		&			\\				
3d$^7$($^4$P)4p						&	u $^5$D$^o$	&	4	&	46720.842	&	213.9697, 215.8919	&	12.2	&			\\				
3d$^7$($^4$P)4p						&	u $^5$D$^o$	&	3	&	46744.993	&	215.7794, 217.1296	&	9.9	&			\\				
3d$^7$($^4$P)4p						&	u $^5$D$^o$	&	2	&	46888.517	&	216.4548, 217.3213	&	11.4	&			\\				
3d$^7$($^4$P)4p						&	u $^5$D$^o$	&	0	&	47171.531	&	215.9923, 341.8507	&	7.8	&			\\				
3d$^7$($^4$P)4p						&	u $^5$D$^o$	&	1	&	47177.234	&	216.3862, 341.7840	&	9.8	&	9.3$^c$		\\\\			
3d$^6$($^3$F$_2$)4s4p($^3$P$^o$)						&	x $^3$F$^o$	&	4	&	46889.142	&	213.2016, 250.1693	&	11.1	&	10.7$^c$		\\				
3d$^6$($^3$F$_2$)4s4p($^3$P$^o$)						&	x $^3$F$^o$	&	3	&	47092.712	&	214.1718, 215.5019	&	18.2	&	17.4$^c$		\\				
3d$^6$($^3$F$_2$)4s4p($^3$P$^o$)						&	x $^3$F$^o$	&	2	&	47197.01	&	215.0184, 215.8734	&	22.9	&	21.8$^c$		\\\\			
3d$^7$($^4$P)4p						&	w $^3$D$^o$	&	3	&	47017.188	&	214.5189, 215.8534	&	11.6	&	11.5$^c$		\\				
3d$^7$($^4$P)4p						&	w $^3$D$^o$	&	2	&	47136.084	&	215.3006, 216.1579	&	11.9	&	11.7$^c$		\\				
3d$^7$($^4$P)4p						&	w $^3$D$^o$	&	1	&	47272.027	&	214.6720, 215.5243	&	14.2	&	15.1$^c$		\\\\				
3d$^6$($^3$P$_2$)4s4p($^3$P$^o$)						&	 $^1$D$^o$	&	2	&	47419.687	&	213.9934, 214.8403	&	33	&			\\\\			
3d$^7$($^2$G)4p						&	z $^1$G$^o$	&	4	&	47452.717 	&	 251.6570, 388.4358	&	66.2	&	59.7$^c$		\\				\\			
3d$^6$($^3$G)4s4p($^3$P$^o$)						&	v $^5$F$^o$	&	5	&	47606.114	&	245.7596, 386.1343	&	27.8	&			\\				
3d$^6$($^3$G)4s4p($^3$P$^o$)						&	v $^5$F$^o$	&	4	&	47929.997	&	246.5149, 248.6691	&	16.8	&			\\				
3d$^6$($^3$G)4s4p($^3$P$^o$)						&	v $^5$F$^o$	&	3	&	48122.928	&	245.3475, 247.4814	&	12.9	&			\\				
3d$^6$($^3$G)4s4p($^3$P$^o$)						&	v $^5$F$^o$	&	2	&	48238.847	&	209.0383, 248.3533	&	11.2	&	11.2$^c$		\\				
3d$^6$($^3$G)4s4p($^3$P$^o$)						&	v $^5$F$^o$	&	1	&	48350.606	&	247.6656, 248.7065	&	11.2	&			\\\\				
3d$^5$($^6$S)4s$^2$4p						&	v $^5$P$^o$	&	3	&	47966.585	&	208.4121, 210.2353	&	6.3	&			\\				
3d$^5$($^6$S)4s$^2$4p						&	v $^5$P$^o$	&	2	&	48163.446	&	209.3684, 210.6394	&	6.4	&			\\				
3d$^5$($^6$S)4s$^2$4p						&	v $^5$P$^o$	&	1	&	48289.871	&	210.0797, 210.8958	&	5.9	&			\\				
						&		&		&		&		&		&			\\				
3d$^7$($^4$P)4p						&	x $^3$P$^o$	&	2	&	48304.643	&	208.7510, 210.8301	&	12.7	&			\\				
3d$^7$($^4$P)4p						&	x $^3$P$^o$	&	1	&	48516.138	&	209.8938, 210.2910	&	10.7	&	10.6$^c$		\\\\				
3d$^7$($^2$G)4p						&	z $^1$H$^o$	&	5	&	48382.603	&	412.0206, 419.9094	&	16.2	&			\\				
						&		&		&		&		&		&			\\				
3d$^7$($^2$G)4p						&	w $^3$F$^o$	&	4	&	49108.896	&	239.5505, 269.2248	&	16.7	&			\\				
3d$^7$($^2$G)4p						&	w $^3$F$^o$	&	3	&	49242.886	&	366.952	&	18.7	&			\\				
3d$^7$($^2$G)4p						&	w $^3$F$^o$	&	2	&	49433.131	&	367.7627	&	12.8	&			\\				
						&		&		&		&		&		&			\\				
3d$^7$($^2$G)4p						&	z $^1$F$^o$	&	3	&	50586.878	&	384.3256	&	18.5	&			\\				
						&		&		&		&		&		&			\\				
3d$^6$($^5$D)4s($^6$D)5p						&	u $^5$F$^o$	&	5	&	51016.66	&	226.7469	&	20.6	&			\\				
3d$^6$($^5$D)4s($^6$D)5p						&	u $^5$F$^o$	&	4	&	51381.457	&	224.8860, 227.1782	&	21.8	&			\\				
						&		&		&		&		&		&			\\				
3d$^6$($^5$D)4s($^6$D)5p						&	t $^5$D$^o$	&	4	&	51076.628	&	226.4389, 228.7631	&	16.4	&			\\				
						&		&		&		&		&		&			\\				
3d$^7$($^2$H)4p						&	u $^3$G$^o$	&	5	&	51373.91	&	253.7458, 337.0783	&	14.5	&	15.4$^c$		\\				
3d$^7$($^2$H)4p						&	u $^3$G$^o$	&	4	&	51668.186	&	311.9494, 336.9547	&	15.2	&	13.5$^c$		\\\\				
3d$^7$($^4$F)5p						&	 $^5$G$^o$	&	6	&	53069.357	&	216.6585	&	55.5	&			\\				
3d$^7$($^4$F)5p						&	 $^5$G$^o$	&	4	&	53852.114	&	216.7386, 238.7282	&	30.3	&			\\				
						&		&		&		&		&		&			\\				
3d$^7$($^4$F)5p						&	 $^5$F$^o$	&	5	&	53084.789	&	216.5861	&	25.5	&			\\				
3d$^7$($^4$F)5p						&	 $^5$F$^o$	&	4	&	53388.637	&	215.1695, 217.2670	&	21.9	&			\\				
						&		&		&		&		&		&			\\				
3d$^7$($^2$H)4p						&	 $^1$I$^o$	&	6	&	53093.529	&	373.8304, 411.8544	&	10.6	&			\\				
						&		&		&		&		&		&			\\				
3d$^7$($^4$F)5p						&	 $^3$F$^o$	&	3	&	54289.034	&	213.0965, 214.7045	&	19.8	&			\\				
						&		&		&		&		&		&			\\				
3d$^6$($^3$D)4s4p($^3$P$^o$)						&	 $^5$D$^o$	&	1	&	53975.744	&	218.1721, 277.3232	&	18.2	&			\\				
3d$^6$($^3$D)4s4p($^3$P$^o$)						&	 $^5$D$^o$	&	4	&	54301.34	&	211.0235, 272.0196	&	19	&			\\				
						&		&		&		&		&		&			\\				
3d$^6$($^3$P$_2$)4s4p($^1$P$^o$)						&	 $^3$D$^o$	&	3	&	57565.305	&	219.2823, 255.1093	&	6.3	&			\\				
\hline		 																										\enddata   							
\tablenotetext{a}{Configurations, terms, level energies, and Ritz wavelengths are from the NIST Atomic Spectra Database (\textit{http://www.nist.gov/pml/data/asd.cfm}).}							\tablenotetext{b}{\cite{ref:obrian1991} TR-LIF with uncertainties equal to the larger of $\pm$5\% or $\pm$0.2~ns.}
\tablenotetext{c}{\cite{ref:langhans1995} TR-LIF with uncertainties of $\pm$10\% for lifetimes $<$3~ns and $\pm$5\% for the remainder.}	
\tablenotetext{d}{\cite{ref:marek1979} delayed coincidence after laser excitation.}
\tablenotetext{e}{\cite{ref:langhans1995}}
																
\label{table:lifetimes}

\end{deluxetable}  

\begin{deluxetable}{crrcc}										
\tablewidth{0pt}										
\tabletypesize{\scriptsize}										
\tablecaption{Completeness of the set of transitions from each upper level estimated by using the calculated branching fractions
of Kurucz (2007).}										
										
\tablehead{										
\colhead{Energy level (cm$^{-1}$)\tablenotemark{a}}		&	&	\colhead{Configuration level\tablenotemark{a}}	&	\colhead{Lifetime used (ns)\tablenotemark{b}} 	&	\colhead{Completeness (\%)\tablenotemark{c}} 	
}										
\startdata										
34843.957		&	&	3d$^7$($^4$F)4p z $^5$G$^o$	&	9.2	&		100	\\
39969.853		&	&	3d$^6$($^5$D)4s4p($^1$P$^o$) x $^5$D$^o$	&	2.7	&		99	\\
40404.518		&	&	3d$^6$($^5$D)4s4p($^1$P$^o$) x $^5$D$^o$	&	2.8	&		99	\\
44022.525		&	&	3d$^6$($^3$F$_2$)4s4p($^3$P$^o$) v $^5$D$^o$	&	96.5	&		99	\\
44415.074		&	&	3d$^6$($^3$F$_2$)4s4p($^3$P$^o$) w $^5$F$^o$	&	41.6	&		99	\\
44760.746		&	&	3d$^6$($^3$F$_2$)4s4p($^3$P$^o$) v $^5$D$^o$	&	21.3	&		98	\\
46720.842		&	&	3d$^7$($^4$P)4p u $^5$D$^o$	&	12.2	&		98	\\
46889.142		&	&	3d$^6$($^3$F$_2$)4s4p($^3$P$^o$) x $^3$F$^o$	&	11.1	&		91	\\
47092.712		&	&	3d$^6$($^3$F$_2$)4s4p($^3$P$^o$) x $^3$F$^o$	&	18.2	&		96	\\
47197.010		&	&	3d$^6$($^3$F$_2$)4s4p($^3$P$^o$) x $^3$F$^o$	&	22.9	&		82	\\
48350.606		&	&	3d$^6$($^3$G)4s4p($^3$P$^o$) v $^5$F$^o$	&	11.2	&		94	\\
48382.603		&	&	3d$^7$($^2$G)4p z $^1$H$^o$	&	16.2	&		99	\\
50586.878		&	&	3d$^7$($^2$G)4p z $^1$F$^o$	&	18.5	&		94	\\
51373.910		&	&	3d$^7$($^2$H)4p u $^3$G$^o$	&	14.5	&		94	\\
53093.529		&	&	3d$^7$($^2$H)4p  $^1$I$^o$	&	10.6	&		98	\\
\enddata										
\tablenotetext{a}{The energy and configuration levels are taken from \cite{ref:kramida2011}.}										
\tablenotetext{b}{Experimental lifetimes measured in this work. The uncertainty of these values is the larger of $\pm$5\% or $\pm$0.2 ns.}	
\tablenotetext{c}{Completeness of the set of transitions from each upper energy level estimated as described in \cite{ref:pickering2001a} and \cite{ref:pickering2001b} by using the calculated branching fractions of \cite{ref:kurucz2007}.}									
\label{table:completeness}										
\end{deluxetable}										

\begin{deluxetable}{ccrcrrcrrccc}																																
\tablewidth{0pt}																																	
\tabletypesize{\scriptsize}																																	
\tablecaption{Experimental BFs, transition probabilities and log$(gf)$ values for 16 odd-parity energy levels of Fe I.}																														
\tablehead{																																	
\colhead{Wavelength\tablenotemark{a}}	&	\multicolumn{2}{c}{Upper Level\tablenotemark{a}}		&	&		\multicolumn{2}{c}{Lower Level\tablenotemark{a}}		&		\colhead{BF\tablenotemark{b}}	&	\colhead{U$_{BF}$\tablenotemark{b}}	&	\colhead{A$_{ul}$\tablenotemark{c}}				&	\colhead{This experiment\tablenotemark{d}}			&		\multicolumn{2}{c}{Published\tablenotemark{e}}								\\
\cline{2-3} \cline{5-6} \cline{11-12}																																	
\colhead{(nm)}	&	\colhead{$E$ (cm$^{-1}$)}	&	\colhead{$J$}	&	&	\colhead{$E$ (cm$^{-1}$)}	&	\colhead{$J$}	&		&	\colhead{(\%)}	&	\colhead{(10$^6$ s$^{-1}$)}				&	\colhead{$\log(gf)$}			&		\colhead{$\log(gf)$}						&	\colhead{Ref.}	
}																																	
\startdata																																	
$358.1193$	&	$34843.957$	&	$6$	&	&	$6928.268$	&	$5$	&	$1.0000$	&	$0.00$	&	$108.70$	(	$5$	)	&	$0.43$	$\pm$	$0.02$	&	$	0.406	$	$\pm$	$	0.005	$	&	BL79	\\
																																	\\
$463.0120$	&	$39969.853$	&	$3$	&	&	$18378.185$	&	$2$	&	$0.0003$	&	$11.9$	&	$0.12$	(	$13$	)	&	$-2.58$	$\pm$	$0.05$	&	$	-2.59	$	$\pm$	$	0.12	$	&	OB91 L	\\
$449.4563$	&	$39969.853$	&	$3$	&	&	$17726.987$	&	$2$	&	$0.0090$	&	$3.4$	&	$3.44$	(	$6$	)	&	$-1.14$	$\pm$	$0.03$	&	$	-1.14	$	$\pm$	$	0.01	$	&	BL82a	\\
$445.9117$	&	$39969.853$	&	$3$	&	&	$17550.180$	&	$3$	&	$0.0065$	&	$3.2$	&	$2.49$	(	$6$	)	&	$-1.29$	$\pm$	$0.03$	&	$	-1.28	$	$\pm$	$	0.01	$	&	BL82a	\\
$312.5651$	&	$39969.853$	&	$3$	&	&	$7985.784$	&	$2$	&	$0.0049$	&	$7.5$	&	$1.90$	(	$9$	)	&	$-1.71$	$\pm$	$0.04$	&	$	-1.66	$	$\pm$	$	0.08	$	&	OB91 L	\\
$310.0665$	&	$39969.853$	&	$3$	&	&	$7728.059$	&	$3$	&	$0.0362$	&	$5.9$	&	$13.94$	(	$8$	)	&	$-0.85$	$\pm$	$0.03$	&	$	-0.87	$	$\pm$	$	0.08	$	&	OB91 L	\\
$306.7244$	&	$39969.853$	&	$3$	&	&	$7376.764$	&	$4$	&	$0.0999$	&	$5.6$	&	$38.42$	(	$8$	)	&	$-0.42$	$\pm$	$0.03$	&	$	-0.51	$	$\pm$	$	0.07	$	&	OB91 L	\\
$254.5978$	&	$39969.853$	&	$3$	&	&	$704.007$	&	$2$	&	$0.1858$	&	$5.1$	&	$71.44$	(	$7$	)	&	$-0.31$	$\pm$	$0.03$	&	$	-0.31	$	$\pm$	$	0.05	$	&	OB91 L	\\
$252.7435$	&	$39969.853$	&	$3$	&	&	$415.933$	&	$3$	&	$0.4774$	&	$3.1$	&	$183.63$	(	$6$	)	&	$0.090$	$\pm$	$0.03$	&	$	0.11	$	$\pm$	$	0.04	$	&	OB91 L	\\
$250.1132$	&	$39969.853$	&	$3$	&	&	$0.000$	&	$4$	&	$0.1794$	&	$5.2$	&	$69.00$	(	$7$	)	&	$-0.34$	$\pm$	$0.03$	&	$	-0.35	$	$\pm$	$	0.05	$	&	OB91 L	\\
																																	\\
$444.7717$	&	$40404.518$	&	$1$	&	&	$17927.381$	&	$1$	&	$0.0138$	&	$4.0$	&	$4.94$	(	$6$	)	&	$-1.36$	$\pm$	$0.03$	&	$	-1.34	$	$\pm$	$	0.01	$	&	BL82a	\\
$440.8414$	&	$40404.518$	&	$1$	&	&	$17726.987$	&	$2$	&	$0.0058$	&	$5.5$	&	$2.08$	(	$8$	)	&	$-1.74$	$\pm$	$0.03$	&	$	-1.78	$	$\pm$	$	0.12	$	&	OB91 L	\\
$309.9895$	&	$40404.518$	&	$1$	&	&	$8154.713$	&	$1$	&	$0.0417$	&	$6.0$	&	$14.91$	(	$8$	)	&	$-1.19$	$\pm$	$0.03$	&	$	-1.08	$	$\pm$	$	0.05	$	&	OB91 L	\\
$308.3741$	&	$40404.518$	&	$1$	&	&	$7985.784$	&	$2$	&	$0.0655$	&	$5.9$	&	$23.41$	(	$8$	)	&	$-1.00$	$\pm$	$0.03$	&	$	-0.88	$	$\pm$	$	0.05	$	&	OB91 L	\\
$253.5607$	&	$40404.518$	&	$1$	&	&	$978.074$	&	$0$	&	$0.2678$	&	$4.7$	&	$95.63$	(	$7$	)	&	$-0.56$	$\pm$	$0.03$	&	$	-0.56	$	$\pm$	$	0.04	$	&	OB91 L	\\
$252.9835$	&	$40404.518$	&	$1$	&	&	$888.132$	&	$1$	&	$0.0881$	&	$6.9$	&	$31.48$	(	$9$	)	&	$-1.04$	$\pm$	$0.04$	&	$	-0.96	$	$\pm$	$	0.04	$	&	OB91 L	\\
$251.8102$	&	$40404.518$	&	$1$	&	&	$704.007$	&	$2$	&	$0.5167$	&	$2.9$	&	$184.54$	(	$6$	)	&	$-0.28$	$\pm$	$0.02$	&	$	-0.26	$	$\pm$	$	0.03	$	&	OB91 L	\\
																																	\\
$377.6455$	&	$44022.525$	&	$4$	&	&	$17550.180$	&	$3$	&	$0.1061$	&	$3.8$	&	$1.10$	(	$6$	)	&	$-1.68$	$\pm$	$0.03$	&	$	-1.49	$	$\pm$	$	0.05	$	&	OB91 L	\\
$275.4427$	&	$44022.525$	&	$4$	&	&	$7728.059$	&	$3$	&	$0.1559$	&	$4.8$	&	$1.62$	(	$7$	)	&	$-1.78$	$\pm$	$0.03$	&	$	-1.78	$	$\pm$	$	0.03	$	&	OB91 L	\\
$272.8021$	&	$44022.525$	&	$4$	&	&	$7376.764$	&	$4$	&	$0.3310$	&	$3.8$	&	$3.43$	(	$6$	)	&	$-1.46$	$\pm$	$0.03$	&	$	-1.46	$	$\pm$	$	0.03	$	&	OB91 L	\\
$269.5036$	&	$44022.525$	&	$4$	&	&	$6928.268$	&	$5$	&	$0.0438$	&	$5.5$	&	$0.45$	(	$8$	)	&	$-2.35$	$\pm$	$0.03$	&	$	-2.33	$	$\pm$	$	0.03	$	&	OB91 L	\\
$229.2525$	&	$44022.525$	&	$4$	&	&	$415.933$	&	$3$	&	$0.3122$	&	$4.0$	&	$3.24$	(	$6$	)	&	$-1.64$	$\pm$	$0.03$	&	$	-1.68	$	$\pm$	$	0.03	$	&	OB91 L	\\
$227.0863$	&	$44022.525$	&	$4$	&	&	$0.000$	&	$4$	&	$0.0484$	&	$6.0$	&	$0.50$	(	$8$	)	&	$-2.46$	$\pm$	$0.03$	&								&		\\
																																	\\
$372.1272$	&	$44415.074$	&	$4$	&	&	$17550.180$	&	$3$	&	$0.0339$	&	$15.3$	&	$0.81$	(	$16$	)	&	$-1.82$	$\pm$	$0.07$	&	$	-1.79	$	$\pm$	$	0.05	$	&	OB91 L	\\
$272.4953$	&	$44415.074$	&	$4$	&	&	$7728.059$	&	$3$	&	$0.1987$	&	$4.6$	&	$4.78$	(	$7$	)	&	$-1.32$	$\pm$	$0.03$	&	$	-1.32	$	$\pm$	$	0.03	$	&	OB91 L	\\
$269.9107$	&	$44415.074$	&	$4$	&	&	$7376.764$	&	$4$	&	$0.2315$	&	$4.4$	&	$5.57$	(	$7$	)	&	$-1.26$	$\pm$	$0.03$	&	$	-1.26	$	$\pm$	$	0.02	$	&	OB91 L	\\
$266.6812$	&	$44415.074$	&	$4$	&	&	$6928.268$	&	$5$	&	$0.3765$	&	$3.6$	&	$9.05$	(	$6$	)	&	$-1.061$	$\pm$	$0.03$	&	$	-1.07	$	$\pm$	$	0.02	$	&	OB91 L	\\
$227.2070$	&	$44415.074$	&	$4$	&	&	$415.933$	&	$3$	&	$0.1170$	&	$5.1$	&	$2.81$	(	$7$	)	&	$-1.71$	$\pm$	$0.03$	&	$	-1.69	$	$\pm$	$	0.02	$	&	OB91 L	\\
$225.0790$	&	$44415.074$	&	$4$	&	&	$0.000$	&	$4$	&	$0.0419$	&	$6.3$	&	$1.01$	(	$8$	)	&	$-2.16$	$\pm$	$0.03$	&	$	-2.08	$	$\pm$	$	0.05	$	&	OB91 L	\\
																																	\\
$273.0982$	&	$44760.746$	&	$1$	&	&	$8154.713$	&	$1$	&	$0.1185$	&	$6.0$	&	$5.56$	(	$8$	)	&	$-1.73$	$\pm$	$0.03$	&	$	-1.68	$	$\pm$	$	0.02	$	&	OB91 L	\\
$271.8436$	&	$44760.746$	&	$1$	&	&	$7985.784$	&	$2$	&	$0.7738$	&	$1.4$	&	$36.33$	(	$5$	)	&	$-0.92$	$\pm$	$0.02$	&	$	-0.90	$	$\pm$	$	0.02	$	&	OB91 L	\\
$228.3304$	&	$44760.746$	&	$1$	&	&	$978.074$	&	$0$	&	$0.0510$	&	$8.3$	&	$2.39$	(	$10$	)	&	$-2.25$	$\pm$	$0.04$	&	$	-2.22	$	$\pm$	$	0.02	$	&	OB91 L	\\
$226.9099$	&	$44760.746$	&	$1$	&	&	$704.007$	&	$2$	&	$0.0383$	&	$10.5$	&	$1.80$	(	$12$	)	&	$-2.38$	$\pm$	$0.05$	&								&		\\
																																	\\
$1033.3185$	&	$46720.842$	&	$4$	&	&	$37045.932$	&	$4$	&	$0.0010$	&	$12.8$	&	$0.09$	(	$14$	)	&	$-1.91$	$\pm$	$0.06$	&								&		\\
$721.9682$	&	$46720.842$	&	$4$	&	&	$32873.630$	&	$4$	&	$0.0065$	&	$6.9$	&	$0.53$	(	$9$	)	&	$-1.43$	$\pm$	$0.04$	&								&		\\
$451.4184$	&	$46720.842$	&	$4$	&	&	$24574.653$	&	$4$	&	$0.0049$	&	$11.2$	&	$0.40$	(	$12$	)	&	$-1.96$	$\pm$	$0.05$	&	$	-1.92	$	$\pm$	$	0.18	$	&	MA74	\\
$435.8501$	&	$46720.842$	&	$4$	&	&	$23783.617$	&	$5$	&	$0.0129$	&	$5.4$	&	$1.06$	(	$7$	)	&	$-1.57$	$\pm$	$0.03$	&	$	-1.68	$	$\pm$	$	0.05	$	&	OB91 P	\\
$399.8052$	&	$46720.842$	&	$4$	&	&	$21715.731$	&	$5$	&	$0.0875$	&	$4.7$	&	$7.17$	(	$7$	)	&	$-0.81$	$\pm$	$0.03$	&	$	-0.91	$	$\pm$	$	0.04	$	&	OB91 P	\\
$383.3308$	&	$46720.842$	&	$4$	&	&	$20641.109$	&	$4$	&	$0.0647$	&	$5.1$	&	$5.30$	(	$7$	)	&	$-0.98$	$\pm$	$0.03$	&	$	-1.032	$	$\pm$	$	0.004	$	&	BL82b	\\
$342.7120$	&	$46720.842$	&	$4$	&	&	$17550.180$	&	$3$	&	$0.6593$	&	$1.9$	&	$54.04$	(	$5$	)	&	$-0.067$	$\pm$	$0.02$	&	$	-0.10	$	$\pm$	$	0.04	$	&	OB91 P	\\
$287.7301$	&	$46720.842$	&	$4$	&	&	$11976.238$	&	$4$	&	$0.0544$	&	$7.2$	&	$4.46$	(	$9$	)	&	$-1.30$	$\pm$	$0.04$	&	$	-1.29	$	$\pm$	$	0.04	$	&	OB91 P	\\
$254.0916$	&	$46720.842$	&	$4$	&	&	$7376.764$	&	$4$	&	$0.0067$	&	$14.0$	&	$0.55$	(	$15$	)	&	$-2.32$	$\pm$	$0.06$	&								&		\\
$251.2275$	&	$46720.842$	&	$4$	&	&	$6928.268$	&	$5$	&	$0.0352$	&	$7.2$	&	$2.89$	(	$9$	)	&	$-1.61$	$\pm$	$0.04$	&	$	-1.73	$	$\pm$	$	0.06	$	&	OB91 P	\\
$215.8920$	&	$46720.842$	&	$4$	&	&	$415.933$	&	$3$	&	$0.0128$	&	$13.4$	&	$1.05$	(	$14$	)	&	$-2.18$	$\pm$	$0.06$	&								&		\\
$213.9698$	&	$46720.842$	&	$4$	&	&	$0.000$	&	$4$	&	$0.0369$	&	$8.7$	&	$3.02$	(	$10$	)	&	$-1.73$	$\pm$	$0.04$	&								&		\\
																																	\\
$553.2747$	&	$46889.142$	&	$4$	&	&	$28819.952$	&	$5$	&	$0.0035$	&	$14.5$	&	$0.31$	(	$15$	)	&	$-1.89$	$\pm$	$0.06$	&	$	-2.10	$	$\pm$	$	0.30	$	&	MA74	\\
$493.4084$	&	$46889.142$	&	$4$	&	&	$26627.607$	&	$4$	&	$0.0014$	&	$17.9$	&	$0.13$	(	$19$	)	&	$-2.39$	$\pm$	$0.07$	&								&		\\
$448.0137$	&	$46889.142$	&	$4$	&	&	$24574.653$	&	$4$	&	$0.0048$	&	$11.7$	&	$0.43$	(	$13$	)	&	$-1.93$	$\pm$	$0.05$	&	$	-1.93	$	$\pm$	$	0.09	$	&	OB91 P	\\
$432.6753$	&	$46889.142$	&	$4$	&	&	$23783.617$	&	$5$	&	$0.0068$	&	$9.8$	&	$0.62$	(	$11$	)	&	$-1.81$	$\pm$	$0.05$	&	$	-1.93	$	$\pm$	$	0.09	$	&	OB91 P	\\
$397.1322$	&	$46889.142$	&	$4$	&	&	$21715.731$	&	$5$	&	$0.0583$	&	$8.6$	&	$5.25$	(	$10$	)	&	$-0.95$	$\pm$	$0.04$	&	$	-0.98	$	$\pm$	$	0.04	$	&	OB91 P	\\
$380.8729$	&	$46889.142$	&	$4$	&	&	$20641.109$	&	$4$	&	$0.0387$	&	$8.9$	&	$3.48$	(	$10$	)	&	$-1.17$	$\pm$	$0.04$	&	$	-1.159	$	$\pm$	$	0.004	$	&	BL82b	\\
$340.7460$	&	$46889.142$	&	$4$	&	&	$17550.180$	&	$3$	&	$0.6629$	&	$2.8$	&	$59.72$	(	$6$	)	&	$-0.029$	$\pm$	$0.02$	&	$	-0.02	$	$\pm$	$	0.04	$	&	OB91 P	\\
$286.3430$	&	$46889.142$	&	$4$	&	&	$11976.238$	&	$4$	&	$0.0462$	&	$7.3$	&	$4.16$	(	$9$	)	&	$-1.34$	$\pm$	$0.04$	&	$	-1.34	$	$\pm$	$	0.04	$	&	OB91 P	\\
$250.1694$	&	$46889.142$	&	$4$	&	&	$6928.268$	&	$5$	&	$0.0524$	&	$7.4$	&	$4.72$	(	$9$	)	&	$-1.40$	$\pm$	$0.04$	&	$	-1.51	$	$\pm$	$	0.05	$	&	OB91 P	\\
$213.2017$	&	$46889.142$	&	$4$	&	&	$0.000$	&	$4$	&	$0.0386$	&	$9.3$	&	$3.48$	(	$11$	)	&	$-1.67$	$\pm$	$0.04$	&	$	-1.33	$	$\pm$	$	0.06	$	&		\\
																												
$730.7931$	&	$47092.712$	&	$3$	&	&	$33412.715$	&	$3$	&	$0.0228$	&	$5.7$	&	$1.26$	(	$8$	)	&	$-1.15$	$\pm$	$0.03$	&	$	-1.53	$	$\pm$	$	0.06	$	&	OB91 P	\\
$435.1544$	&	$47092.712$	&	$3$	&	&	$24118.817$	&	$4$	&	$0.0210$	&	$4.7$	&	$1.15$	(	$7$	)	&	$-1.64$	$\pm$	$0.03$	&	$	-1.73	$	$\pm$	$	0.04	$	&	OB91 P	\\
$412.1802$	&	$47092.712$	&	$3$	&	&	$22838.321$	&	$2$	&	$0.0539$	&	$3.9$	&	$2.96$	(	$6$	)	&	$-1.28$	$\pm$	$0.03$	&	$	-1.45	$	$\pm$	$	0.04	$	&	OB91 P	\\
$398.3956$	&	$47092.712$	&	$3$	&	&	$21999.129$	&	$4$	&	$0.1291$	&	$3.5$	&	$7.09$	(	$6$	)	&	$-0.93$	$\pm$	$0.03$	&	$	-1.02	$	$\pm$	$	0.04	$	&	OB91 P	\\
$383.7135$	&	$47092.712$	&	$3$	&	&	$21038.986$	&	$2$	&	$0.0177$	&	$5.4$	&	$0.97$	(	$7$	)	&	$-1.82$	$\pm$	$0.03$	&	$	-1.78	$	$\pm$	$	0.09	$	&	OB91 P	\\
$381.3058$	&	$47092.712$	&	$3$	&	&	$20874.481$	&	$3$	&	$0.1163$	&	$2.9$	&	$6.39$	(	$6$	)	&	$-1.01$	$\pm$	$0.02$	&	$	-1.07	$	$\pm$	$	0.04	$	&	OB91 P	\\
$377.9416$	&	$47092.712$	&	$3$	&	&	$20641.109$	&	$4$	&	$0.0112$	&	$10.8$	&	$0.61$	(	$12$	)	&	$-2.036$	$\pm$	$0.05$	&	$	-1.99	$	$\pm$	$	0.05	$	&	OB91 P	\\
$340.4354$	&	$47092.712$	&	$3$	&	&	$17726.987$	&	$2$	&	$0.2180$	&	$4.3$	&	$11.98$	(	$7$	)	&	$-0.84$	$\pm$	$0.03$	&	$	-0.88	$	$\pm$	$	0.04	$	&	OB91 P	\\
$338.3979$	&	$47092.712$	&	$3$	&	&	$17550.180$	&	$3$	&	$0.1494$	&	$4.8$	&	$8.21$	(	$7$	)	&	$-1.006$	$\pm$	$0.03$	&	$	-1.11	$	$\pm$	$	0.04	$	&	OB91 P	\\
$292.9618$	&	$47092.712$	&	$3$	&	&	$12968.553$	&	$2$	&	$0.0135$	&	$11.2$	&	$0.74$	(	$12$	)	&	$-2.18$	$\pm$	$0.05$	&	$	-2.22	$	$\pm$	$	0.05	$	&	OB91 P	\\
$289.5035$	&	$47092.712$	&	$3$	&	&	$12560.933$	&	$3$	&	$0.1089$	&	$6.5$	&	$5.98$	(	$8$	)	&	$-1.28$	$\pm$	$0.03$	&	$	-1.43	$	$\pm$	$	0.04	$	&	OB91 P	\\
$284.6830$	&	$47092.712$	&	$3$	&	&	$11976.238$	&	$4$	&	$0.0164$	&	$9.0$	&	$0.90$	(	$10$	)	&	$-2.12$	$\pm$	$0.04$	&	$	-2.13	$	$\pm$	$	0.04	$	&	OB91 P	\\
$253.9587$	&	$47092.712$	&	$3$	&	&	$7728.059$	&	$3$	&	$0.0072$	&	$16.6$	&	$0.40$	(	$17$	)	&	$-2.57$	$\pm$	$0.07$	&								&		\\
$251.7123$	&	$47092.712$	&	$3$	&	&	$7376.764$	&	$4$	&	$0.0321$	&	$7.3$	&	$1.77$	(	$9$	)	&	$-1.93$	$\pm$	$0.04$	&								&		\\
$215.5020$	&	$47092.712$	&	$3$	&	&	$704.007$	&	$2$	&	$0.0157$	&	$25.2$	&	$0.86$	(	$26$	)	&	$-2.38$	$\pm$	$0.10$	&								&		\\
$214.1718$	&	$47092.712$	&	$3$	&	&	$415.933$	&	$3$	&	$0.0212$	&	$15.3$	&	$1.16$	(	$16$	)	&	$-2.25$	$\pm$	$0.07$	&								&		\\
																																	\\
$744.3022$	&	$47197.010$	&	$2$	&	&	$33765.304$	&	$2$	&	$0.0125$	&	$13.0$	&	$0.55$	(	$14$	)	&	$-1.64$	$\pm$	$0.06$	&								&		\\
$437.3561$	&	$47197.010$	&	$2$	&	&	$24338.765$	&	$3$	&	$0.0230$	&	$6.7$	&	$1.01$	(	$8$	)	&	$-1.84$	$\pm$	$0.04$	&	$	-1.83	$	$\pm$	$	0.09	$	&	OB91 P	\\
$437.2987$	&	$47197.010$	&	$2$	&	&	$24335.764$	&	$2$	&	$0.0064$	&	$22.6$	&	$0.28$	(	$23$	)	&	$-2.40$	$\pm$	$0.09$	&	$	-2.58	$	$\pm$	$	0.18	$	&	MA74	\\
$412.2516$	&	$47197.010$	&	$2$	&	&	$22946.814$	&	$1$	&	$0.0710$	&	$5.7$	&	$3.10$	(	$8$	)	&	$-1.40$	$\pm$	$0.03$	&	$	-1.39	$	$\pm$	$	0.04	$	&	OB91 P	\\
$410.4154$	&	$47197.010$	&	$2$	&	&	$22838.321$	&	$2$	&	$0.0073$	&	$8.8$	&	$0.32$	(	$10$	)	&	$-2.40$	$\pm$	$0.04$	&								&		\\
$382.1835$	&	$47197.010$	&	$2$	&	&	$21038.986$	&	$2$	&	$0.1889$	&	$4.7$	&	$8.25$	(	$7$	)	&	$-1.044$	$\pm$	$0.03$	&	$	-1.10	$	$\pm$	$	0.04	$	&	OB91 P	\\
$341.5531$	&	$47197.010$	&	$2$	&	&	$17927.381$	&	$1$	&	$0.0839$	&	$6.6$	&	$3.67$	(	$8$	)	&	$-1.49$	$\pm$	$0.04$	&	$	-1.39	$	$\pm$	$	0.05	$	&	OB91 P	\\
$339.2305$	&	$47197.010$	&	$2$	&	&	$17726.987$	&	$2$	&	$0.1783$	&	$5.8$	&	$7.79$	(	$8$	)	&	$-1.17$	$\pm$	$0.03$	&	$	-1.07	$	$\pm$	$	0.05	$	&	OB91 P	\\
$292.0691$	&	$47197.010$	&	$2$	&	&	$12968.553$	&	$2$	&	$0.1060$	&	$7.2$	&	$4.63$	(	$9$	)	&	$-1.53$	$\pm$	$0.04$	&	$	-1.39	$	$\pm$	$	0.04	$	&	OB91 P	\\
$288.6317$	&	$47197.010$	&	$2$	&	&	$12560.933$	&	$3$	&	$0.0335$	&	$10.8$	&	$1.47$	(	$12$	)	&	$-2.039$	$\pm$	$0.05$	&	$	-2.09	$	$\pm$	$	0.04	$	&	OB91 P	\\
$256.0557$	&	$47197.010$	&	$2$	&	&	$8154.713$	&	$1$	&	$0.0405$	&	$7.7$	&	$1.77$	(	$9$	)	&	$-2.061$	$\pm$	$0.04$	&	$	-2.11	$	$\pm$	$	0.04	$	&	OB91 P	\\
$254.9525$	&	$47197.010$	&	$2$	&	&	$7985.784$	&	$2$	&	$0.0081$	&	$21.3$	&	$0.35$	(	$22$	)	&	$-2.77$	$\pm$	$0.09$	&	$	-2.49	$	$\pm$	$	0.05	$	&	OB91 P	\\
$253.2876$	&	$47197.010$	&	$2$	&	&	$7728.059$	&	$3$	&	$0.0328$	&	$8.0$	&	$1.43$	(	$10$	)	&	$-2.16$	$\pm$	$0.04$	&	$	-2.16	$	$\pm$	$	0.04	$	&	OB91 P	\\
$215.0185$	&	$47197.010$	&	$2$	&	&	$704.007$	&	$2$	&	$0.0311$	&	$18.2$	&	$1.36$	(	$19$	)	&	$-2.33$	$\pm$	$0.08$	&								&		\\
																														
$393.5307$	&	$48350.606$	&	$1$	&	&	$22946.814$	&	$1$	&	$0.0102$	&	$16.0$	&	$0.91$	(	$17$	)	&	$-2.199$	$\pm$	$0.07$	&	$	-1.82	$	$\pm$	$	0.18	$	&	MA74	\\
$328.6016$	&	$48350.606$	&	$1$	&	&	$17927.381$	&	$1$	&	$0.0106$	&	$20.6$	&	$0.94$	(	$21$	)	&	$-2.34$	$\pm$	$0.08$	&								&		\\
$326.4513$	&	$48350.606$	&	$1$	&	&	$17726.987$	&	$2$	&	$0.0261$	&	$15.2$	&	$2.33$	(	$16$	)	&	$-1.95$	$\pm$	$0.06$	&	$	-1.32	$	$\pm$	$	0.05	$	&	OB91 P	\\
$248.7066$	&	$48350.606$	&	$1$	&	&	$8154.713$	&	$1$	&	$0.6014$	&	$2.6$	&	$53.70$	(	$6$	)	&	$-0.83$	$\pm$	$0.02$	&	$	-0.75	$	$\pm$	$	0.05	$	&	OB91 P	\\
$247.6657$	&	$48350.606$	&	$1$	&	&	$7985.784$	&	$2$	&	$0.2956$	&	$4.9$	&	$26.39$	(	$7$	)	&	$-1.14$	$\pm$	$0.03$	&	$	-1.08	$	$\pm$	$	0.04	$	&	OB91 P	\\
																																	\\
$537.9574$	&	$48382.603$	&	$5$	&	&	$29798.934$	&	$4$	&	$0.0128$	&	$3.9$	&	$0.79$	(	$6$	)	&	$-1.42$	$\pm$	$0.03$	&		$-1.51$	$\pm$	$0.04$						&	OB91 P	\\
$524.2491$	&	$48382.603$	&	$5$	&	&	$29313.006$	&	$6$	&	$0.0526$	&	$2.7$	&	$3.25$	(	$6$	)	&	$-0.83$	$\pm$	$0.02$	&	$	-0.97	$	$\pm$	$	0.04	$	&	OB91 P	\\
$511.0358$	&	$48382.603$	&	$5$	&	&	$28819.952$	&	$5$	&	$0.0148$	&	$4.0$	&	$0.91$	(	$6$	)	&	$-1.41$	$\pm$	$0.03$	&	$	-1.37	$	$\pm$	$	0.04	$	&	OB91 P	\\
$459.5358$	&	$48382.603$	&	$5$	&	&	$26627.607$	&	$4$	&	$0.0087$	&	$5.0$	&	$0.54$	(	$7$	)	&	$-1.73$	$\pm$	$0.03$	&	$	-1.76	$	$\pm$	$	0.04	$	&	OB91 P	\\
$419.9095$	&	$48382.603$	&	$5$	&	&	$24574.653$	&	$4$	&	$0.8193$	&	$0.6$	&	$50.57$	(	$5$	)	&	$0.17$	$\pm$	$0.02$	&	$	0.16	$	$\pm$	$	0.04	$	&	OB91 P	\\
$412.0206$	&	$48382.603$	&	$5$	&	&	$24118.817$	&	$4$	&	$0.0345$	&	$3.1$	&	$2.13$	(	$6$	)	&	$-1.23$	$\pm$	$0.03$	&	$	-1.27	$	$\pm$	$	0.04	$	&	OB91 P	\\
$378.9176$	&	$48382.603$	&	$5$	&	&	$21999.129$	&	$4$	&	$0.0317$	&	$4.1$	&	$1.96$	(	$7$	)	&	$-1.33$	$\pm$	$0.03$	&	$	-1.29	$	$\pm$	$	0.04	$	&	OB91 P	\\
$360.3681$	&	$48382.603$	&	$5$	&	&	$20641.109$	&	$4$	&	$0.0060$	&	$11.7$	&	$0.37$	(	$13$	)	&	$-2.10$	$\pm$	$0.05$	&	$	-2.01	$	$\pm$	$	0.08	$	&	OB91 P	\\
$347.5863$	&	$48382.603$	&	$5$	&	&	$19621.005$	&	$5$	&	$0.0115$	&	$18.4$	&	$0.71$	(	$19$	)	&	$-1.85$	$\pm$	$0.08$	&							&		\\\\

$470.8969$	&	$50586.878$	&	$3$	&	&	$29356.742$	&	$2$	&	$0.0068$	&	$28.4$	&	$0.37$	(	$29$	)	&	$-2.07$	$\pm$	$0.11$	&	$	-2.03	$	$\pm$	$	0.09	$	&	OB91 P	\\
$454.7847$	&	$50586.878$	&	$3$	&	&	$28604.611$	&	$2$	&	$0.1279$	&	$2.1$	&	$6.91$	(	$6$	)	&	$-0.82$	$\pm$	$0.02$	&	$	-1.01	$	$\pm$	$	0.12	$	&	OB91 P	\\
$417.1900$	&	$50586.878$	&	$3$	&	&	$26623.733$	&	$2$	&	$0.0221$	&	$6.1$	&	$1.20$	(	$8$	)	&	$-1.66$	$\pm$	$0.03$	&	$	-1.70	$	$\pm$	$	0.05	$	&	OB91 P	\\
$410.3611$	&	$50586.878$	&	$3$	&	&	$26224.967$	&	$3$	&	$0.0022$	&	$18.8$	&	$0.12$	(	$20$	)	&	$-2.67$	$\pm$	$0.08$	&								&		\\
$384.3257$	&	$50586.878$	&	$3$	&	&	$24574.653$	&	$4$	&	$0.7374$	&	$0.8$	&	$39.86$	(	$5$	)	&	$-0.21$	$\pm$	$0.02$	&	$	-0.24	$	$\pm$	$	0.04	$	&	OB91 P	\\
$380.8282$	&	$50586.878$	&	$3$	&	&	$24335.764$	&	$2$	&	$0.0177$	&	$16.3$	&	$0.95$	(	$17$	)	&	$-1.84$	$\pm$	$0.07$	&	$	-1.94	$	$\pm$	$	0.06	$	&	OB91 P	\\
$352.7891$	&	$50586.878$	&	$3$	&	&	$22249.428$	&	$3$	&	$0.0148$	&	$21.3$	&	$0.80$	(	$22$	)	&	$-1.98$	$\pm$	$0.09$	&								&		\\
$302.6056$	&	$50586.878$	&	$3$	&	&	$17550.180$	&	$3$	&	$0.0161$	&	$26.2$	&	$0.87$	(	$27$	)	&	$-2.077$	$\pm$	$0.10$	&								&		\\
																																	\\
$697.7429$	&	$51373.910$	&	$5$	&	&	$37045.932$	&	$4$	&	$0.0068$	&	$10.7$	&	$0.47$	(	$12$	)	&	$-1.42$	$\pm$	$0.05$	&								&		\\
$540.3822$	&	$51373.910$	&	$5$	&	&	$32873.630$	&	$4$	&	$0.0319$	&	$3.1$	&	$2.20$	(	$6$	)	&	$-0.98$	$\pm$	$0.03$	&	$	-1.03	$	$\pm$	$	0.05	$	&	OB91 P	\\
$453.1636$	&	$51373.910$	&	$5$	&	&	$29313.006$	&	$6$	&	$0.0048$	&	$12.1$	&	$0.33$	(	$13$	)	&	$-1.95$	$\pm$	$0.05$	&								&		\\
$443.2568$	&	$51373.910$	&	$5$	&	&	$28819.952$	&	$5$	&	$0.0117$	&	$12.0$	&	$0.80$	(	$13$	)	&	$-1.58$	$\pm$	$0.05$	&	$	-1.56	$	$\pm$	$	0.18	$	&	MA74	\\
$395.6455$	&	$51373.910$	&	$5$	&	&	$26105.906$	&	$6$	&	$0.2287$	&	$2.5$	&	$15.77$	(	$6$	)	&	$-0.39$	$\pm$	$0.02$	&	$	-0.34	$	$\pm$	$	0.04	$	&	OB91 P	\\
$373.0386$	&	$51373.910$	&	$5$	&	&	$24574.653$	&	$4$	&	$0.1296$	&	$4.9$	&	$8.94$	(	$7$	)	&	$-0.69$	$\pm$	$0.03$	&	$	-0.65	$	$\pm$	$	0.04	$	&	OB91 P	\\
$337.0783$	&	$51373.910$	&	$5$	&	&	$21715.731$	&	$5$	&	$0.3413$	&	$3.9$	&	$23.54$	(	$6$	)	&	$-0.36$	$\pm$	$0.03$	&	$	-0.27	$	$\pm$	$	0.04	$	&	OB91 P	\\
$325.2915$	&	$51373.910$	&	$5$	&	&	$20641.109$	&	$4$	&	$0.0267$	&	$9.8$	&	$1.84$	(	$11$	)	&	$-1.49$	$\pm$	$0.05$	&	$	-1.42	$	$\pm$	$	0.04	$	&	OB91 P	\\
$312.5683$	&	$51373.910$	&	$5$	&	&	$19390.167$	&	$6$	&	$0.1049$	&	$5.7$	&	$7.23$	(	$8$	)	&	$-0.93$	$\pm$	$0.03$	&	$	-0.87	$	$\pm$	$	0.04	$	&	OB91 P	\\
$253.7459$	&	$51373.910$	&	$5$	&	&	$11976.238$	&	$4$	&	$0.0604$	&	$9.6$	&	$4.17$	(	$11$	)	&	$-1.35$	$\pm$	$0.05$	&	$	-1.47	$	$\pm$	$	0.06	$	&	OB91 P	\\
																																	\\
$420.3938$	&	$53093.528$	&	$6$	&	&	$29313.006$	&	$6$	&	$0.0342$	&	$7.2$	&	$3.23$	(	$9$	)	&	$-0.95$	$\pm$	$0.04$	&	$	-0.99	$	$\pm$	$	0.04	$	&	OB91 P	\\
$411.8545$	&	$53093.528$	&	$6$	&	&	$28819.952$	&	$5$	&	$0.5906$	&	$3.2$	&	$55.72$	(	$6$	)	&	$0.27$	$\pm$	$0.03$	&	$	0.22	$	$\pm$	$	0.04	$	&	OB91 P	\\
$373.8305$	&	$53093.528$	&	$6$	&	&	$26351.038$	&	$5$	&	$0.3624$	&	$5.2$	&	$34.18$	(	$7$	)	&	$-0.031$	$\pm$	$0.03$	&	$	-0.03	$	$\pm$	$	0.04	$	&	OB91 P	\
\enddata																																	
\tablenotetext{a}{Wavelengths, upper and lower energy levels and J quantum numbers are taken from \cite{ref:kramida2011}.}																																	
\tablenotetext{b}{The measured branching fraction, BF, is expressed per-unit and its relative uncertainty, $\delta$BF/BF, as a percentage. }																																	
\tablenotetext{c}{The measured transition probability, A$_{ul}$, in 10$^6$ s$^{-1}$. In brackets, its uncertainty expressed in percentage.}																																	
\tablenotetext{d}{The log$(gf)$ values measured in this work together with their uncertainty in dex.}																																	
\tablenotetext{e}{Values of log$(gf)$s from other authors used for comparison with their uncertainty in dex. The acronyms in the reference column correspond to: BL79 - Blackwell et al. (1979); OB91 - O'Brian et al. (1991); BL82a - Blackwell et al. (1982a); MA74 - May et al. (1974); BL82b - Blackwell et al. (1982b); BA94 - Bard et al. (1994). The letter included after the reference OB91 indicates the method used by the authors, with `L' and `P' standing for `lifetime' and `population' method, respectively.} 																																	
\label{table:results}																																	
\end{deluxetable}																																	

\begin{deluxetable}{rrrrrrlrrcrrrrr}																																
\tablewidth{0pt}																																	
\tabletypesize{\scriptsize}																																	
\tablecaption{Lines from Table~\ref{table:results} selected for solar
synthesis.}																														
\tablehead{																																	
\colhead{$\lambda_{air}$}	&\colhead{$E_{\rm low}$}	&\colhead{VdW\tablenotemark{a}}&\multicolumn{2}{c}{This experiment} && \multicolumn{3}{c}{Previously published} & \colhead{$v_{\rm macro}$} & \multicolumn{2}{c}{New gf} && \multicolumn{2}{c}{Previous $gf$}	\\
\cline{4-5} \cline{7-9}\cline{11-12} \cline{14-15}																																	
\colhead{(\AA)} & \colhead{(eV)} & \colhead{parameter} &   \colhead{log($gf$)} & \colhead{Unc.} && \colhead{log($gf$)} & \colhead{Unc.\tablenotemark{b}} & \colhead{Ref.\tablenotemark{c}} & \colhead{(km~s$^{-1}$)} & \colhead{log($\varepsilon$)} & \colhead{RMS\tablenotemark{d}} && \colhead{log($\varepsilon$)} & \colhead{RMS\tablenotemark{d}}	
}																						
									
\startdata			

3808.729 & 2.559 &  265.262 & $-$1.17 & 0.04 && $-$1.16 & 0.00 & BL82b & 3.2 & 7.47 & 0.9 && 7.46 & 0.9 \\
4120.206 & 2.990 &  338.253 & $-$1.23 & 0.03 && $-$1.27 & 0.04 &  OB91 & 3.6 & 7.56 & 1.9 && 7.60 & 2.0 \\
4432.568 & 3.573 &  275.254 & $-$1.58 & 0.05 && $-$1.56 & 0.18 &  MA74 & 3.2 & 7.39 & 1.0 && 7.38 & 1.0 \\
4447.717 & 2.223 &  429.302 & $-$1.36 & 0.03 && $-$1.34 & 0.01 & BL82a & 3.2 & 7.60 & 0.6 && 7.58 & 0.7 \\
4459.117 & 2.176 &  417.302 & $-$1.29 & 0.03 && $-$1.28 & 0.01 & BL82a & 2.5 & 7.50 & 0.9 && 7.49 & 0.9 \\
4494.563 & 2.198 &  416.302 & $-$1.14 & 0.03 && $-$1.14 & 0.01 & BL82a & 3.1 & 7.46 & 1.2 && 7.46 & 1.2 \\
4514.184 & 3.047 &  296.271 & $-$1.96 & 0.05 && $-$1.92 & 0.18 &  MA74 & 3.2 & 7.40 & 1.1 && 7.36 & 1.2 \\
4547.847 & 3.546 &  313.266 & $-$0.82 & 0.02 && $-$1.01 & 0.12 &  OB91 & 3.1 & 7.41 & 1.5 && 7.59 & 1.6 \\
4595.358 & 3.301 &  286.270 & $-$1.73 & 0.03 && $-$1.76 & 0.04 &  OB91 & 3.4 & 7.60 & 1.2 && 7.62 & 1.2 \\
4630.120 & 2.279 &  416.254 & $-$2.58 & 0.05 && $-$2.59 & 0.12 &  OB91 & 3.0 & 7.56 & 1.7 && 7.56 & 1.7 \\
4708.969 & 3.640 &   $-$7.800 & $-$2.07 & 0.11 && $-$2.03 & 0.09 &  OB91 & 4.1 & 7.92 & 0.7 && 7.88 & 0.7 \\
5379.574 & 3.695 &  363.249 & $-$1.42 & 0.03 && $-$1.51 & 0.04 &  OB91 & 3.2 & 7.43 & 0.9 && 7.53 & 0.9 \\
5403.822 & 4.076 &   $-$7.810 & $-$0.98 & 0.03 && $-$1.03 & 0.05 &  OB91 & 3.7 & 7.62 & 1.1 && 7.66 & 1.1 \\
5532.747 & 3.573 &  237.255 & $-$1.89 & 0.06 && $-$2.10 & 0.30 &  MA74 & 3.4 & 7.25 & 0.4 && 7.45 & 0.4 \\
7219.682 & 4.076 &   $-$7.740 & $-$1.43 & 0.04 && $-$1.73 &      &   K14 & 3.1 & 7.40 & 0.5 && 7.70 & 0.5 \\
7307.931 & 4.143 &   $-$7.810 & $-$1.15 & 0.03 && $-$1.53 & 0.06 &  OB91 & 3.4 & 7.33 & 0.4 && 7.69 & 0.5 \\
7443.022 & 4.186 &   $-$7.810 & $-$1.64 & 0.06 && $-$1.40 &      &   K14 & 3.4 & 7.47 & 0.5 && 7.24 & 0.5 \\

\enddata

\tablenotetext{a}{Van der Waals broadening parameter (see text for
explanation).} 
\tablenotetext{b}{Uncertainties are only available
for experimentally measured log($gf$) values.}
\tablenotetext{c}{Reference acronyms are the same as for
Table~\ref{table:results}. In addition, K14 stands for \cite{ref:kurucz2014}. 
}
\tablenotetext{d}{RMS difference between observed and synthetic flux in percent, for the points included in the fit.}
\label{tab:solar}
\end{deluxetable}

\end{document}